\documentclass[12pt, letterpaper, oneside]{article}

\usepackage{graphicx}
\usepackage[body={6.5in, 9in}, right=1in, top=1in]{geometry}
\usepackage{amssymb}
\usepackage{amsmath} 

\newcommand\ee{\end{equation}}
\newcommand\be{\begin{equation}}
\newcommand\eea{\end{eqnarray}}
\newcommand\bea{\begin{eqnarray}}
\newcommand{\sfrac}[2]{{\textstyle\frac{#1}{#2}}}
\newcommand\di{\partial}

\begin{document}

\begin{center}
\LARGE{\textbf{The quantum mechanics of perfect fluids}} \\[1cm]
\large{Solomon Endlich$^{\rm a}$, Alberto Nicolis$^{\rm a}$,  Riccardo Rattazzi$^{\rm b}$, 
and Junpu Wang$^{\rm a}$}
\\[0.4cm]

\vspace{.2cm}
\small{\textit{$^{\rm a}$ Department of Physics and ISCAP, \\ 
Columbia University, New York, NY 10027, USA}}

\vspace{.2cm}
\small{\textit{$^{\rm b}$ Institut de Th\'eorie des Ph\'enom\`enes Physiques, EPFL,\\
CH1015 Lausanne, Switzerland}}

\end{center}

\vspace{.2cm}

\begin{abstract}
We consider the canonical quantization of an ordinary fluid. The resulting long-distance effective field theory is derivatively coupled, and therefore strongly coupled in the UV. The system however exhibits a number of peculiarities, associated with the vortex degrees of freedom. On the one hand, these have formally a {\em vanishing} strong-coupling energy scale, thus suggesting that the effective theory's  regime of validity is vanishingly narrow. On the other hand, we prove an analog of Coleman's theorem, whereby the semiclassical vacuum has no quantum counterpart, thus suggesting that the vortex premature strong-coupling phenomenon stems from a bad identification of the ground state and of the perturbative degrees of freedom. Finally, vortices break the usual connection between short distances and high energies, thus potentially impairing the unitarity of the effective theory.
\end{abstract}

\section{Introduction}

Empirically, all fluids we know of undergo a phase transition when we lower the temperature. Either they freeze, or they transform into more exotic systems, like  super-fluids or Fermi liquids.
Why are there no {\em ordinary} fluids at zero temperature? For weakly coupled systems we have a microscopic understanding of the low-temperature macroscopic behavior at finite density \cite{LL} (we also understand why such systems at high temperatures exhibit hydrodynamic behavior.)
However for strongly coupled ones, such as the so-called non-Fermi liquids, we don't. Of course a strongly coupled system could look like anything at long-distances---there is no in-principle preference for the ordinary fluid dynamics. Nevertheless, classical hydrodynamics is so common in nature at high temperatures, and as we will see, it can be defined purely in terms of low-energy degrees of freedom and symmetries like an ordinary QFT, that it is natural to ask whether there exist strongly-coupled quantum systems that at zero temperature and finite density do behave like ordinary fluids.

It is tempting to conjecture that we know of no such systems simply because the corresponding quantum effective field theory would be inconsistent. As usual, the advantage of this viewpoint is that, as long as we allow for the most generic local dynamics involving all long-distance degrees of freedom and compatible with the symmetries, we are allowed to be completely agnostic about the microphysics yielding such macroscopic dynamics.
Further motivation to investigate the consistency of the ordinary-fluid effective theory comes from the recent interest in the so-called holographic liquids at low temperatures. There, one deals with strongly coupled systems at finite density via a dual description in terms of classical gravity, which can be thought of as providing {\em the} microphysics behind these liquid-like states. Characterizing the long-distance dynamics of such systems is non-trivial however (see e.g.~\cite{NS} and references therein).
Our approach allows us to make progress on the (perhaps modest) question: can some of these low-temperature, finite-density systems behave like ordinary fluids? Partial indication 
that the answer may be `yes' comes from the results of \cite{EJL}, but \cite{NS} argues that these cannot be consistently interpreted as a sign of hydrodynamic behavior.

Without committing to any models for the microphysics, we will argue that: {\em (i)} the effective theory of an ordinary fluid is not consistent; {\em (ii)} the effective theory of an ordinary fluid might be consistent
after all; {\em (iii)} there is no guarantee that such effective theory be unitary at low energies. Clearly, the matter deserves further study. 

Before starting our quantitative analysis which substantiates these claims, we conclude this introductory section with a few qualifications.
First, we will only consider fluids at zero temperature. Thus, our results cannot be readily exported to the finite-temperature case. Indeed we know that at high temperatures ordinary fluids abound in the real world (hence the `ordinary'), and they exhibit no funny quantum effects at long distances like those we are going to discuss. In these cases we expect quantum effects to be overwhelmed by thermal ones. It would be interesting to validate this expectation quantitatively---a task which we leave for future work.

Second, we are going to neglect dissipative effects throughout our paper. That is, we will consider {\em perfect}  fluids only.  The reason is that dissipative effects in hydrodynamics, like for instance those parameterized by viscosity and heat conduction, are associated with higher derivative corrections to the perfect fluid dynamics---see e.g.~\cite{Weinberg}. Therefore in the far infrared, that is for processes taking place on sufficiently long length scales, they can be safely ignored. Moreover, one may expect that the actual coefficients weighing these higher derivative corrections approach zero when the temperature is taken to zero. For instance in \cite{NS} it is argued that a finite viscosity at zero temperature is incompatible with hydrodynamic behavior. So, it is conceivable that by working at low enough temperatures and at long enough distances, one can make dissipation doubly negligible.

Third, a crucial role in our analysis will be played by vortices. Precisely the existence of `light' vortices is what distinguishes an ordinary fluid from a superfluid at the classical level. As long as one concentrates on the compressional modes---the sound waves---both systems obey hydrodynamics \cite{LL}, and this holds at the non-linear, relativistic level as well \cite{DGNR}.
However in a superfluid the velocity field is irrotational, which implies that any vortex-like configuration will be singular at the center of the vortex, along a line, with the curl of the velocity field behaving like a delta-function peaked on this line. This means that from the viewpoint of the long distance/low energy effective field theory, the vortices are really UV-degrees of freedom, with finite energy per unit length (which is in fact mildly IR-divergent.) So, for instance, one cannot form vortices by scattering phonons of very low energies---there is a gap, and as long as one works below the gap, the vortex degrees of freedom can be ignored
\footnote{A close relative of the superfluid vortex is the roton excitation, which is also gapped, and which can then also be neglected in the far infrared.}.
On the other hand, in ordinary fluids, one can build vortex configurations that are arbitrarily mild, that is, that involve arbitrarily low momenta only. As a consequence, there is no gap in the energy one can store in a vortex. Vortices in an ordinary fluid are low-energy degrees of freedom, and they belong in the low-energy/long distance effective field theory together with the sound waves.
In fact, we will see that, in a sense to be made precise below, the vanishing of the vortex gap is stronger than that of the sound wave gap---vortices are `more massless' than sound waves.
This will be the origin of all the quantum-mechanical peculiarities we will discuss, which are therefore absent for a superfluid.

\section{The classical theory}

To begin with, let us review how classical hydrodynamics can be cast into a field theoretical language. We will adopt the viewpoint and notation of \cite{DGNR}, to which we refer the reader for details.
In particular, we will parameterize the fluid's configuration space by giving at time $t$ the comoving (`Lagrangian') coordinates $\phi^I$ of each fluid element, as a function of the physical (`Eulerian') position $\vec x$  occupied by that fluid element:
\be
\phi^I = \phi^I(\vec x, t) \; , \qquad I=1,2,3 \; . 
\ee
Of course this description is completely equivalent to the inverse one, whereby one gives $\vec x$ as a function of the comoving coordinates and of time, and which we will also use at some point. However we find our starting point more convenient to construct the theory, because it identifies the fluid's macroscopic degrees of freedom with three scalar functions of spacetime coordinates, i.e.~with three scalar fields: this way keeping track of Poincar\'e invariance is straigthforward, and so is coupling the fluid to other systems, like gravity for instance. Moreover, as we will see in a moment, hydrodynamics follows straightforwardly via standard effective field theory (EFT) logic once we identify the correct internal symmetries.

Before proceeding, it is worth stressing that we will be dealing with a fully relativistic theory, even though most laboratory fluids are highly non-relativistic. 
For these one could impose Galilean invariance rather than Poincar\'e invariance, but this would not simplify the analysis we are going to carry out, conceptually or algebraically. We thus see no reason why not to keep track of relativistic effects and just neglect them when appropriate.

We now come to the symmetries. The spacetime ones are of course the Poincar\'e group, under which our $\phi^I$'s transform as scalars. As for the internal ones, we have a huge redundancy in choosing the fluid's comoving coordinates. This is not a symmetry---it is the standard arbitrariness one has in parameterizing a Lagrangian system's configuration space. To make any progress, we should make an explicit choice. A particularly convenient one is the following: At some given reference pressure we demand that for the homogeneous and static fluid configuration---the fluid's `ground state'---the comoving coordinates be aligned with the physical ones:
\be  \label{classical_vacuum}
\phi^I = x^I \; .
\ee
It is then clear that homogeneity and isotropy for the physical properties of such a state cannot emerge unless the dynamics are invariant under {\em internal} translations and rotations:
\bea 
\label{shifts} \phi^I & \to & \phi^I + a^I \\ 
\label{rot} \phi^I & \to & O^I {}_J \, \phi^J \; ,
\eea
where $O$ is an $SO(3)$ matrix.
So far, we have not specified what distinguishes a fluid from an isotropic solid (a `jelly'). It is an additional symmetry---the invariance under volume-preserving diffeomorphisms---
\be \label{diff}
\phi^I \to \xi^I (\phi^J) \; , \qquad \det \frac{\di \xi^I}{\di \phi^J} = 1 \; .
\ee
This should not be confused with a trivial relabeling of the comoving coordinates, which we already got rid of. Rather, it corresponds to an invariance of the dynamics under physically moving fluid elements around without compressing or dilating the fluid anywhere. If we were to do so in a solid, we would feel transverse stresses trying to pull all volume elements back to their rest position. In a fluid, on the other hand, we only feel reaction forces against compression or dilation.

Invariance under shifts (eq.~(\ref{shifts})) forces each field $\phi^I$ to be acted upon by at least one-derivative. At low momenta/low frequencies, the most relevant terms are those with the fewest derivatives. Therefore, the lowest order low-energy Lagrangian will involve exactly one derivative acting on each $\phi^I$. Poincar\'e invariance then forces the Lagrangian to depend on the matrix
\be
B^{IJ} = \di_\mu \phi^I \, \di^\mu \phi^J 
\ee
only. Internal rotations (eq.~(\ref{rot})) impose that we focus on $SO(3)$ invariant functions of $B^{IJ}$, and the volume preserving diffs (eq.~(\ref{diff})) select the determinant among these. We thus have that the most generic low-energy Lagrangian compatible with all the symmetries is \cite{DGNR}
\be \label{action}
S = \int \! d^4 x \, F(B) \; , \qquad B \equiv \det B^{IJ} \; ,
\ee
where $F$ is a generic function.

It is straightforward to check that the action (\ref{action}) describes the dynamics of a perfect fluid. The stress energy tensor is
\be
T_{\mu\nu} = -2 F'(B) B \, B^{-1}_{IJ} \, \di_\mu \phi^I \di_\nu \phi^J + \eta_{\mu\nu} F(B)
\ee
(we are using the `mostly plus' signature for the metric), which matches the standard form $T_{\mu\nu} = (\rho + p) u_\mu u_\nu + p \,\eta_{\mu\nu}$ upon the identifications \cite{DGNR}
\be \label{rho_p_u}
\rho = -F(B) \; , \qquad p = F(B) - 2 F'(B) B \; , \qquad u^\mu = \frac{1}{6\sqrt B} \epsilon^{\mu \alpha \beta \gamma} \epsilon_{IJK} \,  \di_\alpha \phi^I \di_\beta \phi^J \di_\gamma \phi^K \; .
\ee
In particular, we see that both $\rho$ and $p$ depend just on the degree of compression $B$, or equivalently, $p$ depends on $\rho$ only---our fluid is `barotropic'. Different choices for $F(B)$ thus correspond to different equations of state $p(\rho)$, and once the equation of state is given, $F(B)$ is uniquely determined. Notice that $\rho$, $p$, and $u^\mu$ are all invariant under our internal symmetries, eqs.~(\ref{shifts}--\ref{diff}), and so is $T_{\mu\nu}$. In fact, $u^\mu$ is invariant under {\em generic} internal diffs, with no volume-preserving restriction. What matters for characterizing the fluid flow is just that comoving coordinates do not change along it. Such a  requirement is clearly preserved by generic diffeomorphisms of the 
comoving coordinates.

Since we have the correct stress-energy tensor for a fluid, we also have the correct hydrodynamical equations, which follow from stress-energy conservation. The classical ground state of the fluid---the equilibrium configuration at a given pressure or density---is given by eq.~(\ref{classical_vacuum}). This spontaneously breaks all of our spacetime and internal symmetries, except for the diagonal combinations of internal shifts and spacial translations, and of internal rotations and spacial ones.
As a result, there are gapless Goldstone bosons---the phonons---of which only the longitudinal one propagates. Indeed, we can study the propagation of small perturbations of the ground state  by splitting $\phi^I = x^I + \pi^I$ and expanding the action at second order in the $\pi$'s. We get \cite{DGNR}
\be \label{free_action}
S_2 = \int \! d^4 x \,  (-F'(1)) \big[ \sfrac12 \dot {\vec \pi} \, ^2 - \sfrac12 c_s ^2 \, \big( \vec \nabla \cdot \vec \pi \big)^2 \big] 
\ee
where we defined the coefficient $c_s^2$ as
\be \label{cs2}
c^2_s = \frac{2 F''(B) B+F'(B)}{F'(B)} \bigg|_{B=1} 
\ee
and we stopped differentianting between internal indices and spacial ones, since they transform in the same way under the unbroken combination of internal rotations and spacial ones. In other words, from now on we should think of $\vec \pi$ as a spacial vector field. We see from the quadratic action for $\vec \pi$ that only its longitudinal component  has a gradient energy. The corresponding free solutions are plane waves propagating with speed $c_s$---the speed of sound. From the expressions for $\rho$ and $p$ as a function of $B$, eq.~(\ref{rho_p_u}), one realizes that $c_s^2 = dp / d \rho \big| _{B=1}$, thus making contact with the usual expression for the sound speed in a perfect fluid. For a non-relativistc fluid $c^2_s \ll 1$, whereas for an ultra-relativistic one $c_s^2 \simeq 1/3$. We will not commit to either case, but  instead leave $c^2_s$ as a generic parameter.

The transverse excitations do not have a gradient energy and as a consequence obey a free particle-like equation of motion, whose general solution is linear in time:
\be
\vec \pi_T = \vec \nabla \times \big( \vec a(\vec x) + \vec b(\vec x ) \cdot t \big) \; ,
\ee
where $\vec a$ and $\vec b$ are arbitrary vector functions.
This is the linearized limit of a vortex in constant rotation. For this reason we will refer to the transverse excitations as `vortices'. Their lack of gradient energy is, of course, a direct consequence of the volume-preserving internal diff invariance, eq.~(\ref{diff}), and is at the origin of all the peculiarities we are going to unveil. Notice, however, that our diff-invariance is not a local symmetry, and as a consequence the configurations spanned by it---the vortices---are not gauge-modes, but real dynamical degrees of freedom. For instance, they have non-vanishing conjugate momenta; they just do not feature wave solutions.

A more complete analysis of this classical field theory is carried out in \cite{DGNR}, where a number of non-trivial results are derived---most notably a relativistic generalization of Kelvin's theorem, and the equivalence between the zero-vorticity sector of our fluid and a superfluid (i.e.~a derivatively coupled scalar with a time-dependent background.) Here, instead, we will consider the quantum theory, and try to make sense of it.

\section{The naive effective theory}\label{eft}

The structure of the quadratic Lagrangian (\ref{free_action}) already signals that, upon canonical quantization, we might be facing a strong-coupling problem for the vortices. The reason is the following: Consider first as a toy model a  quantum-mechanical oscillator with some anharmonic corrections to the potential. In perturbation theory, one first solves the harmonic problem, thus getting the standard oscillator spectrum, and then treats the anharmonicities as small corrections. The approximation is justified for those  states whose wavefunctions are localized in a region where the potential is dominated by its quadratic approximation. So, for perturbation theory to be applicable in this case, one needs at least the ground state to have a localized enough  wave-function (highly excited states will always be outside the regime of validity of perturbation theory.) Of course, what localizes the  ground state is the curvature of the harmonic potential---the oscillator's frequency. For the system to be `weakly coupled', one thus needs a steep enough quadratic potential. If we now move on to field theory, the role of the quadratic potential is usually played---in the absence of mass terms---by the gradient energy. For given spatial momentum $\vec k$, the gradient energy gives a potential $\propto k^2 |\varphi_k|^2$. The vacuum wavefunction is thus localized about $\varphi_k = 0$, and cannot probe large field values where interactions may become important. In the absence of a gradient energy, on the other hand, each mode's vacuum wavefunction is totally delocalized in the quadratic approximation, and its dynamics are completely determined by the interactions. We thus reach the conclusion that a (massless) field theory without gradient energies is prone to strong coupling, at all scales.

There is a number of caveats in applying the above logic to our case. The first is that the absence of gradient energy may be an accidental feature of the lowest order in the derivative expansion. This is the case, for instance, for the ghost condensate \cite{ghost}, where gradient energy starts at the four-derivative order, $E_{\rm grad}\propto \big(\nabla^2 \pi \big)^2$. In the absence of quadratic terms with fewer spatial derivatives, such a term cannot be relegated to the class of higher-dimension operators, because it is marginal by definition---together with the kinetic energy $E_{\rm kin} \propto \dot \pi^2$ it determines how things behave under rescalings. In this case then, there is a well defined perturbative expansion. But this way out is not available to our vortices: the absence of gradient energy for them is enforced by a symmetry, which also forbids higher spatial-derivative quadratic terms. In the absence of time-dependence, exciting vortices costs nothing: we can deform the ground state $\phi^I = x^I$ in the `transverse' direction via eq.~(\ref{diff}) and pay no energy price, and this extends to non-linear order as well. 
The second caveat, more relevant for us, is that the above quantum oscillator toy model assumes that the anharmonic interactions are of the potential form---only in this case delocalization of the wavefunction {\em necessarily} leads to strong coupling, because having access to large values of $q$ entails having access to large interactions. 
But in our case, by construction, we only have derivative interactions, and moreover the very same symmetry that forbids the vortex gradient energy is also going to forbid many interactions involving vortices. In particular, as we will see more concretely in the following, all vortex interactions that do not involve at least two time derivatives are forbidden. 
Therefore the connection between wavefunction delocalization and strong-coupling is less obvious in our case. 

To settle the question, we should probe the theory by computing some physical quantity and check whether the perturbative expansion holds. The ideal candidates are usually $S$-matrix elements, but here we face a complication.
The longitudinal phonon has standard wave solutions, which upon canonical quantization, get mapped onto standard free-particle states.
The transverse phonons, in contrast, do not behave as waves, and as a consequence there are no quantum asymptotic states associated with them. The classical field $\vec \pi_T$ behaves like a collection of infinitely many free particles rather than infinitely many oscillators. Upon quantization, its Hilbert space is not made up of standard Fock states. Without asymptotic states there is no $S$-matrix.

A possible alternative, is to compute instead local $n$-point functions in real space, and to check whether perturbation theory holds for them. They may be as physical as the $S$-matrix: they characterize the physical interaction among local  sources that couple to our fluid. We do not need asymptotic states to set up such a question. For instance, we can define the theory and the associated correlation functions via the path-integral formulation.
Another possibility, which we will choose, is to give the theory asymptotic states for the vortex degrees of freedom by deforming it in the IR. 
We can add to the classical action a term that is compatible with all the symmetries except for the volume-preserving diffs,
\be
\Delta S = F'(1)  \int \! d^4 x \,   \sfrac12 c_T^2  \, B^{II} \; , \qquad c_T^2 \ll c_s^2
\ee
and whose only effect, once expanded about the ground state, is to introduce a small gradient energy for the transverse Goldstones \footnote{More precisely, the expansion of $B^{II}$ is
\be
B^{II}= - \dot{\vec \pi} \, ^2 + 2 \, \vec \nabla \cdot \vec \pi + \big( \nabla_i \pi^j \, \nabla_i \pi^j  \big) \; .
\ee
The linear term is a total derivative, and can thus be neglected. The other terms, on top of giving the transverse phonons a gradient energy, correct the kinetic and gradient energies already present in (\ref{free_action}). However in the limit $c_T^2 \ll c_s^2 < 1$ these corrections are also negligible.
}:
\be
S_2 \to \int \! d^4 x \,  (-F'(1)) \big[ \sfrac12 \dot {\vec \pi} \, ^2 - \sfrac12 c_s ^2 \, \big( \vec \nabla \cdot \vec \pi_L \big)^2 - \sfrac12 c_T ^2 \, \big( \nabla_i \pi_T^j \, \nabla_i \pi_T^j  \big)\big]  \; .
\ee

We thus have wave solutions, propagating with speed $c_T$, for the `vortices' in the deformed theory---we promoted the vortices to real transverse phonons. Essentially, we are deforming the fluid into a solid/jelly that is stiff under compressional stresses but very soft under transverse ones. The original theory is recovered in the $c_T \to 0$ limit, with a qualification of course. With this $c_T$ deformation we are perturbing drastically the far infrared of the theory. We are adding asymptotic states, and we are going from not having an $S$-matrix to having one. So from this viewpoint the fluid limit is obviously discontinuous. However, we expect more local quantities like $n$-point functions to be continuous in this limit. The situation should be similar to having a fairly narrow unstable particle: strictly speaking it is not an asymptotic state, yet for processes happening at time- and distance-scales much shorter than the particle's lifetime, we can treat it as an asymptotic state and associate an $S$-matrix to it. For scattering processes faster than $\sim 10$ minutes, neutrons behave like asymptotic states.

So, concretely, here is our program. We will consider scattering and decay processes in the $c_T$-deformed theory. In particular, for simplicity we will stick to processes that involve at most four external legs. Thus to carry out calculations at tree-level, we need to expand the action up to quartic order in the $\vec \pi$ field; this is done in the Appendix, and the result is reported below.
By construction, interactions involve one derivative per field. For finite $c_T$, the theory is a standard derivatively coupled theory, and thus strongly-coupled in the UV. The strong coupling scale will depend on all parameters of the theory; however, we are interested in the $c_T$-dependence, since eventually we will be taking $c_T$ to zero while keeping everything else fixed. If the strong coupling energy scale slides to zero in this limit, or equivalently, if cross sections and decay rates at fixed momentum or energy blow up in this limit, the theory is strongly-coupled at all scales, and thus inconsistent.
Notice that we are trying to ascertain the consistency of the theory by computing something---the $S$-matrix---that loses its meaning in the limit we are interested in. Still  at finite $c_T$ we expect that the strong-coupling scale for the $S$-matrix be related to a similar strong-coupling scale for $n$-point functions---that is, that for the latter perturbation theory break down at a distance-scale given by the strong-coupling scale inferred from the $S$-matrix. Thus, even though the $S$-matrix does not exist in the $c_T \to 0$ limit, the formal fact that it appears to be strongly coupled at all scales is probably signaling that real-space $n$-point functions cannot be reliably computed in the fluid theory, at any distance scales.

It is convenient to rewrite the original Lagrangian (\ref{action}) as
\be \label{f}
{\cal L} =  -  w_0 f \big( \sqrt B \big)
\ee
where $w_0 = - 2 F'(1) = (\rho + p)_{B=1}$ is the ground-state's enthalpy density, and $f$ is normalized accordingly, so that $f'(1)=1$. With this new notation the speed of sound (\ref{cs2}) is simply $c_s^2 = f''(1)/f'(1) = f''(1)$. Note that the derivatives here are with respect to $\sqrt{B}$.
Also, we will use $\di \pi$ to denote the matrix with entries $(\di \pi)_{ij} = \di_i \pi_j$, and the brackets $[ \,  \dots \, ]$ to denote the trace of the matrix within.
Then, up to fourth order the action is (see the Appendix)
\bea
{\cal L} & \to &  w_0 \Big\{ \sfrac12 \dot {\vec \pi}^2 - \sfrac12 c_s^2 [\di \pi]^2  - \sfrac12 c_T^2 [\di \pi^T \di \pi ]\nonumber \\
&& + \, \sfrac12 c_s^2 [\di \pi] [\di \pi^2] - \sfrac16 \big( 3 c_s^2 + f_3 \big) [\di \pi]^3 +    \sfrac12(1+c_s^2) \,  [\di \pi] \dot {\vec \pi}^2  - \dot {\vec \pi} \cdot \di \pi \cdot \dot {\vec \pi} \nonumber \\
& & - \, c_s^2 [\di \pi] \det \di \pi   - \sfrac18 c_s^2   [\di \pi^2]^2   + \sfrac14  \big( c_s^2 + f_3 \big) [\di \pi^2][\di \pi]^2  
-\sfrac1{24} \big( 3 c_s^2 +  6 f_3 +f_4 \big) [\di \pi]^4  \nonumber \\
  && + \, \dot {\vec \pi} \cdot \di \pi^2 \cdot \dot {\vec \pi}  
- (1+c_s^2) [\di \pi] \, \dot {\vec \pi} \cdot \di \pi \cdot \dot {\vec \pi} + \sfrac12 | \di \pi^T \cdot \dot {\vec \pi}|^2 \nonumber \\
 && +    \,  \sfrac14 \big( (1+3 c_s^2 + f_3) \, [\di \pi]^2 - (1+c_s^2) \,  [\di \pi^2] \big) \dot {\vec \pi}^2  + \sfrac18 (1- c_s^2) \, \dot {\vec \pi}^4 
  \Big\} \; . 			\label{full_action}
\eea
The first line is the free part of the Lagrangian, including the $c_T$-deformation. The second line collects the trilinear interactions, whereas the third and fourth lines collect the quartic ones. $f_3$ and $f_4$ stand for $f'''(1)$ and $f''''(1)$, respectively. Finally, notice that via the suffix $T$ we indicate the transpose of a matrix, rather than the transverse part of $\vec \pi$ as we did above.

At this order we have four free parameters: $c_s^2$, $c_T^2$, $f_3$, and $f_4$. The dimensionful overall factor of $w_0$ just gives us some reference units---we could use units in which it is one. For $c_T^2$, we know that we want $c_T^2 \ll c_s^2$. As to $c_s^2$, it will be much smaller than one for a non-relativistic fluid, and of order one ($1/3$) for an ultra-relativistic one. In the former case we expect $f_3$ and $f_4$ to be naturally of order $c_s^2$, or smaller. If they were larger, $B=1$ would be a special point for the shape of $f(\sqrt B)$, since by going to, say, $B=2$, the second derivative of $f$, which controls $c_s^2$, would undergo a relative change of more than order one. Likewise, in the ultra-relativistic case, for the same reason we probably want $f_3$ and $f_4$ of order one, or smaller. That is, if we assume that $B=1$ is a fairly generic point for $f$, $f_3$ and $f_4$ have to be at most of order $c_s^2$.~\footnote{This is what happens for instance
for the constant (i.e., $B$-independent) $c_s^2$ Lagrangian
\be
{\cal L} = -w_0 \big( \sqrt{B} \big) ^{1+c_s^2} \; ,
\ee
which corresponds to the simple equation of state $p = c_s^2 \rho$. Notice however that such a simple case, besides being extensively considered by cosmologists, is not preferred in any sense over more generic equations of state---with the exception of the ultra-relativistic case, where the linear equation of state $p=1/3 \, \rho$ follows from scale-invariance.}
On the other hand it may be possible to have a fluid with some feature in the equation of state, where $f''$ is small but higher derivatives of $f$ are large.
In the following we will make no assumptions about these couplings, since even carefully chosen values for them do not lead to drastic simplifications for our computations. Similarly, the non-relativistic case $c_s^2 \ll 1$ is only slightly simpler that the fully relativistic one, and we thus see no reason why not to investigate the latter.



\subsection{Sound-wave strong-coupling scale}\label{strong_coupling}

As a warmup, we estimate the strong-coupling scale for longitudinal phonon scatterings by ignoring the vortices, both as external states as well as internal lines. We assume for simplicity that $f_3$ and $f_4$ are of their `natural' size, $c_s^2$. We also assume that there are no cancellations among the various interactions. 
So, schematically the structure of the action is
\be
S_{\rm sound} \sim \int d^3 x  \, dt \, w_0 \big[ (\dot \pi^2 - c_s^2 \,  \di^2 \pi^2) + c_s^2 \, \di^3 \pi^3 + c_s^2  \, \di^4 \pi^4 \big] \; ,
\ee
where $\di$ stands for a typical spatial derivative, and for the interactions we used $\dot \pi \sim c_s \, \di \pi$, valid for not terribly off-shell phonons.
Now, to estimate the size of the amplitude at a given energy or momentum, we can proceed as follows. First, we redefine the time variable
\be
t \to t / c_s \; ,
\ee
to get a relativistic-looking kinetic term:
\bea
S_{\rm sound} & \sim & \int d^3 x  \, dt /c_s \, w_0 \big[ c_s^2(\dot \pi^2 -  \di^2 \pi^2) + c_s^2 \, \di^3 \pi^3 + c_s^2 \,  \di^4 \pi^4 \big] \\ 
&\sim & w_0 c_s  \int d^4 x  \, \big[ (\dot \pi^2 -  \di^2 \pi^2 ) +  \di^3 \pi^3  + \di^4 \pi^4  \big]  \; .
\eea
Then, we notice that  $c_s$ has factored out of the action, and that combined with $w_0$ it gives the only energy/momentum scale in the action: $M^4 \equiv w_0 c_s$. The rest has standard relativistic scaling (without Lorentz-invariant contractions though), which means that we can apply standard relativistic amplitude estimates. All interactions inside the integral have unit coefficient; the typical $2 \to 2$ dimensionless amplitude is thus $k^4$, combined with the appropriate powers of $M$ to match dimensionality
\be \label{estimate_amplitude}
\mbox{interaction strength} \sim \frac{k^4}{w_0 c_s} \sim \frac{E^4}{w_0 c_s^5} \; ,
\ee
where $E$ is the typical energy in units of the original time variable (notice that spatial coordinates are untouched, so there is no such ambiguity for $k$.)
The strong-coupling momentum and energy are thus
\be
k_* = (w_0 c_s)^{1/4} \; , \qquad E_* = c_s \, k_* \; ,
\label{strongs}
\ee
respectively.

The above estimate yields the correct strong-coupling scale for longitudinal phonons.
We could do the same for the vortex sector, at finite $c_T$. However as we will see, for the vortex interactions there are cancellations that are not manifest in eq.~(\ref{full_action}) and that would impair this simple estimate.

\subsection{Hunting for all factors of $c$}

When we start computing amplitudes and physical quantities like cross sections and decay rates, we have to be careful about extra factors of $c_s$ and of $c_T$ besides those appearing explicitly in the various interaction terms. For instance, we just saw that the longitudinal phonon interaction strength is of order $k^4/(w_0 c_s)$, whereas the Lagrangian interaction terms are proportional to $c_s^2$. In hindsight, this result just follows from dimensional analysis, once we keep separate units for space and time (we can still set $\hbar=1$ though.) The quartic interactions involve four powers of momentum. The interaction strength is thus $k^4$ divided by whatever combination of $w_0$ and $c_s$ has the same units as $k^4$. Of course, we have an ambiguity as to the units of $w_0$---is it a mass- or energy-density? We can easily resolve this ambiguity by looking at the kinetic term.  By construction our $\pi^I$ have units of length; the action is dimensionless (for $\hbar=1$); $w_0$
is thus a mass density, and $w_0 c_s^2$ an energy density: $[\rho_0 c_s^2] = E k^3 = c_s k^4$. The dimensionless combination therefore is $c_s k^4 /(\rho_0 c_s^2) = k^4 /(\rho_0 c_s)$, as expected.

So, a possible strategy to get all the factors of $c_s$ right in amplitudes, cross sections, and rates, is to use the standard relativistic formulae, and then insert suitable powers of $c_s$ to match dimensions. Essentially this is equivalent to redefining the time variable as we did above, to end up with a relativistic kinetic term with $c_s=1$. However this strategy is going to fail once we include vortices/transverse phonons in our processes: with two different propagation speeds $c_s$ and $c_T$ for longitudinal and transverse polarizations, dimensional analysis does not suffice. Equivalently, by redefining the time variable we can cast only one of the two kinetic terms in relativistic form.

In the Appendix we will therefore briefly review the standard relativistic formulae and derive the modifications needed to apply them to our case with $c \neq 1$.
 We adopt this somewhat cumbersome action plan, rather than going through some standard condensed matter textbook and trying to dig up the relevant non-relativistc formulae, for no other reason than we are more familiar with the relativistic Feynman rules and related formulae---and we assume that the reader is also.
The bottom line is pleasantly surprising: We can use the standard relativistic Feynman rules and formulae for infinitesimal cross-sections and rates, with no modifications, even when we start considering different fields with different speeds. By `standard relativistic rules and formulae' we mean those associated with the so-called relativistic normalization of single-particle states, as derived for instance in  Peskin-Shroeder \cite{Peskin}.

As to the overall factor of $w_0$ in (\ref{full_action}), it is straightforwardly kept track of. Either by inserting for each Feynman diagram a $w_0$ for each vertex, a  $1/w_0$ for each internal line, and a  $1/\sqrt{w_0}$ for each external line, or most simply by setting it to one and retrieving it at the end of the computation via dimensional analysis.

\section{Simple processes -- the vortex strong coupling}\label{processes}

We compute, to tree level, a number of simple processes in order of increasing number of vortices on the external legs. 
For scattering processes, for simplicity we will only consider initial states with zero total momentum. Given that Lorentz boosts are spontaneously broken and that we have a preferred reference frame, this is a non-trivial choice---we are setting some kinematic invariants to zero. With an abuse of language, we will refer to this choice as ``working
in the center of mass (CM) frame.'' For the decay of a single finite energy excitation, on the other hand, this choice is not an option, of course.
We will use the formul\ae~for amplitudes, cross sections and rates found in the Appendix. But, as commented on before, the good news is that these formul\ae~look just like the usual relativistic ones that we are used to dealing with. So, except for the additional factors of $c_s$ and $c_T$ coming from the external states' dispersion relations and from the internal lines' propagators, everything goes through just as usual: each external line carries a polarization-vector factor (times $1/\sqrt{w_0}$), each incoming or outgoing time-derivative contributes a $\mp i \omega$, each incoming or outgoing spacial-gradient contributes a $\pm i \vec k$, and so on. From eq.~(\ref{full_action}), we immediately get the Feynman propagator:
\be
\langle T \pi^I(x) \pi^J(y) \rangle \to \frac{1}{w_0} \cdot \frac{i P_L^{IJ}}{\omega^2 - c_s^2 p^2 + i \epsilon} + \frac{1}{w_0} \cdot \frac{i P_T^{IJ}}{\omega^2 - c_T^2 p^2 +i\epsilon} \; ,
\ee
where $P_L$ and $P_T$ are the longitudinal and transverse projectors, respectively.

We will not content ourselves with amplitudes. Rather, we will compute physical, measurable quantities like cross sections and decay rates. The reason is that amplitudes depend crucially on the normalization chosen for the single-particle states. For instance going from the so-called relativistic normalization to the non-relativistic one, would move some factors of $c_s$ and $c_T$ from the amplitudes to the phase-space elements, in such a way as to keep cross-sections and rates unaffected. Ascertaining the strong-coupling of the theory in the $c_T \to 0$ limit at the level of amplitudes requires a derivation of partial waves, a la  Jacob-Wick, being careful about the factors of $c_s$ and $c_T$. Although we have also derived our results using that method, we found it simpler to present them by focussing on cross-sections and decay rates.

A final remark about external vortices. When we take the $c_T \to 0$ limit we have to decide whether we are going to keep their momenta or their energies fixed. The first choice is the more conservative, since it corresponds to taking their energies to zero, thus weakening any possible strong-coupling phenomenon we are going to encounter. It is also the only consistent one, since the alternative one would send the vortex momenta to infinity, outside the regime of validity of any effective theory. In the following we parameterize everything in terms of momenta rather than energies, so that taking the $c_T\to 0$ limit is straightforward.
Notice also that only if we keep the vortex momenta fixed is our deformed theory with small $c_T$ close to the fluid one with $c_T=0$: in the Lagrangian $c_T^2$ weighs the gradient energy, so that by sending $c_T$ to zero while keeping the momenta fixed one is in fact sending the magnitude of that Lagrangian term to  zero.
Related to this, it is somewhat tricky to deal with processes that include longitudinal phonons in the initial state but no longitudinal phonon in the final state: the initial longitudinal phonons' finite energy should be divided among the final vortices, thus making their momenta diverge for $c_T \to 0$. In other words, one cannot send $c_T$ to zero and keep all momenta fixed. We will see an example of this below, in the decay of a longitudinal phonon into two transverse vortices. We postpone a discussion of the related subtleties until then.

We will use $\vec p\,$'s to denote the momenta of the longitudinal modes, and $\vec k$'s and $\hat \epsilon$'s to denote the momenta and polarizations of the transverse modes. Our $\hat \epsilon$'s are real, thus corresponding to linear polarizations, and normalized to one (hence the `hat'.) For longitudinal phonons the polarization vector is $\hat p$, of course. For all the processes we will just compute the leading contribution in  the limit $c_T/c_s\ll 1$, for which we hope to learn something about the original fluid ($c_T=0$).


\subsection{Longitudinal $2 \to  2$ scattering}\label{LLtoLL}

This is the simplest of the scattering processes.  To tree level, the only relevant diagrams are:
\begin{center}
\includegraphics[width=0.7\textwidth]{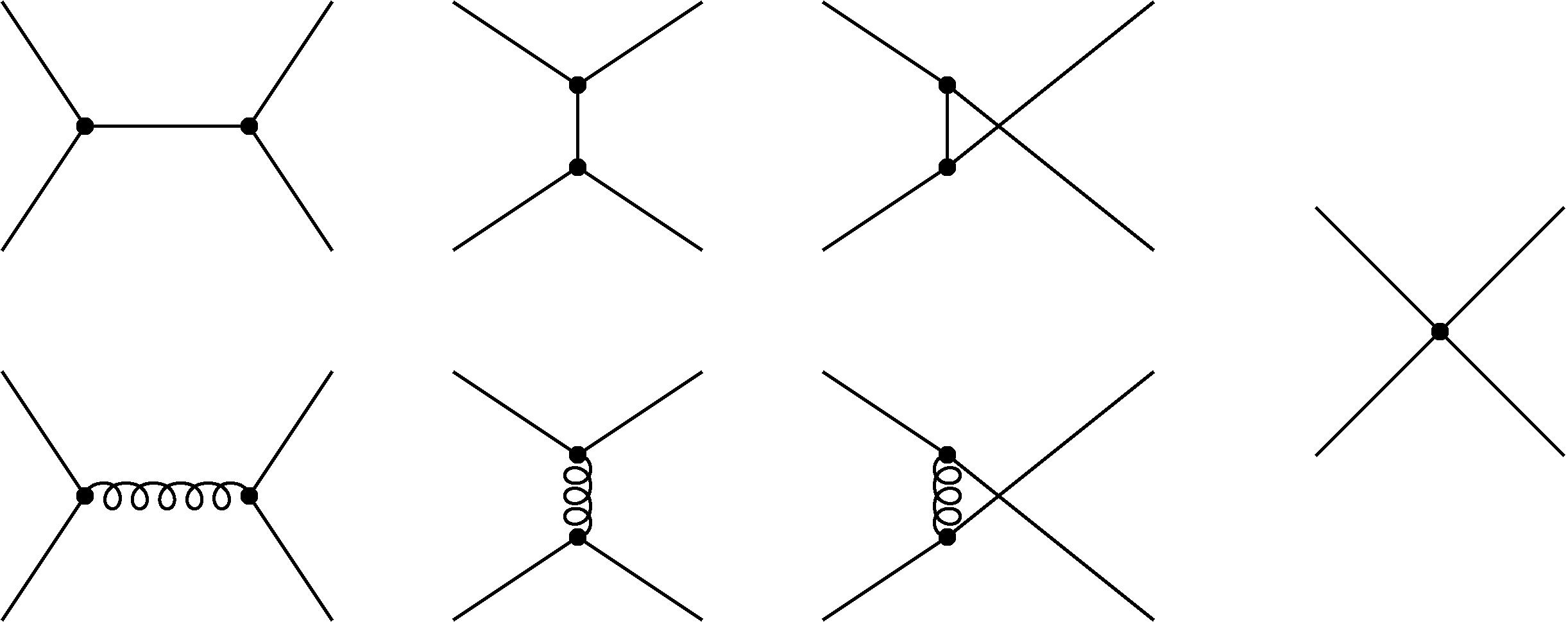}
\end{center}
We designate, here and for the rest of the paper, the solid lines as longitudinal excitations and the curly lines as the transverse excitations. Time flows to the right. 

When done in the center of mass frame, the only kinematic variables are the momentum of the longitudinal phonons $p$ and the scattering angle $\theta$. To tree level, the total amplitude is given by
\begin{equation}
i \mathcal{M}_{LL\rightarrow LL}= -\frac{i p^4 c_s^2}{w_0}\big[ f_4/c_s^2  -2f_3^2/c_s^4  + 3 c_s^2 + 2 f_3 + c_s^4 + 2 (1-3 c_s^2) \cos^2 \theta    \big]
\end{equation}
Remarkably, the graphs with transverse propagators do not contribute to the amplitude, even individually. The infinitesimal cross section is
\begin{equation}
d\sigma=\frac{1}{ c_s^6 }\frac{|\mathcal{M}_{LL\rightarrow LL}|^2}{64 \pi^2 (2p)^2} d\Omega \; ,
\end{equation}
where we made use of the phase space element computed in the Appendix (eq.~(\ref{dPi_zeroP})).
We can easily calculate the total cross section. The final particles are identical, so we over-count when we integrate over all final phase space. To counteract this we simply include a $1/2$ symmetry factor. To all orders in $c_s$ the total cross-section is
\begin{equation} \label{sigmaL}
\sigma_{LL\rightarrow LL}=\frac{1}{256 \pi}\frac{1}{p^2}\left(\frac{p^4}{w_0c_s}\right)^2\left[2 \alpha^2+\frac{4 \alpha \beta}{3}+\frac{2 \beta^2}{5}\right] \sim \frac{1}{p^2}\left(\frac{p^4}{w_0c_s}\right)^2
\end{equation}
where $\alpha\equiv ( f_4/c_s^2  -2f_3^2/c_s^4  + 3 c_s^2 + 2 f_3 + c_s^4 )= \mathcal{O}(1)+\mathcal{O}(c_s^2)+\mathcal{O}(c_s^4)$ (assuming $f_3, f_4 \sim c_s^2$) and $\beta \equiv 2(1-3c_s^2)$.

This result matches our dimensional estimate of the strong coupling scale for longitudinal phonons  in eq.~(\ref{strongs}): the cross-section (\ref{sigmaL}) is the geometric cross-sectional area for wave-packets of wavelength $1/p$, times the square of the dimensionless interaction strength we estimated in sect.~\ref{strong_coupling}. We have strong coupling when $\sigma$ becomes of order $1/p^2$---in such a case the two wave-packets have an  ${\cal O}(1)$ probability of interacting---in agreement with sect.~\ref{strong_coupling}.


\subsection{Longitudinal decay and vorticity production}

In addition to scattering cross sections, we can also calculate the decay rate of a longitudinal phonon into a longitudinal phonon and a transverse one. This is kinematically allowed thanks to the difference in propagation speeds between longitudinal and transverse excitations. To tree level, we simply have one diagram, the longitudinal-longitudinal-transverse vertex:
\begin{center}
\includegraphics[width=0.2\textwidth]{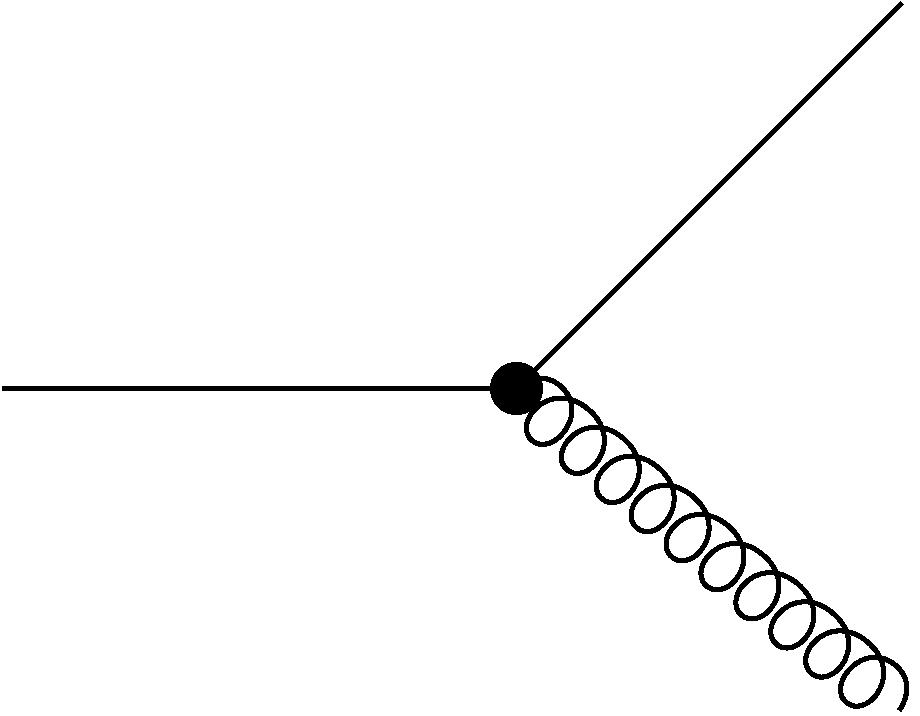}
\end{center}
Imposing the kinematical constraints coming from momentum and energy conservation, expanding in $c_T/c_s$, and keeping only to first order in this parameter, we can write the amplitude as
\begin{equation}
i\mathcal{M}_{L\rightarrow LT}=\frac{-4 c_T c_s \, p^3} {\sqrt{w_0}} (\hat{\epsilon} \cdot \hat p) \sin \theta/2 \, (\cos \theta - c_s^2) 
\label{LLT}
\end{equation}
where  $\theta$ is the angle between the outgoing longitudinal mode and the decaying one (with $\theta = 0$ corresponding to forward decay.)
If the vortex is polarized orthogonally to the scattering plane the decay amplitude vanishes because of parity conservation, whereas for parallel polarization it further simplifies to
\be
i\mathcal{M}_{L\rightarrow LT_\parallel}=\frac{-2 c_T c_s \, p^3} {\sqrt{w_0}}  \sin \theta \, (\cos \theta - c_s^2) 
\ee
Notice the fairly featureless angular dependence, which combined with a similarly featureless phase space (eq.~(\ref{dPi_nonzeroP})) yields the surprising result that with an order-one relative probability the longitudinal phonon will recoil {\em backwards}, i.e.~with $\theta>\pi/2$, by emitting a high-momentum forward vortex.
Moreover, the unusual kinematical constraints associated with the $c_T \ll c_s$ hierarchy force the vortex to be emitted always in the forward half-space, and at an obtuse angle relative to the final longitudinal phonon.

The physical quantity that we want to calculate is the decay rate $\Gamma$, given by (\ref{decay rate}). Using the infinitesimal phase space for a two-particle final state with non-zero total momentum and $c_T \ll c_s$, given by (\ref{dPi_nonzeroP}), summing over possible transverse polarizations and integrating over the solid angle we get:
\begin{equation} \label{GammaLtoLT}
\Gamma_{L \to LT}=\frac{c_T}{c_s}\frac{p^5}{w_0} \, \frac{4}{315\pi}(5-6c_s^2+21c_s^4)
\end{equation}
which, as we can see, smoothly goes to zero as $c_T \to 0$.

In the same way we can study the decay $L\to TT$ of a longitudinal phonon in  two vortices. Notice that the decay kinematics in the $c_T/c_s\ll 1$ limit implies $k_1\simeq -k_2$ and $2 |k_1|\simeq |p| c_s/c_T\gg |p|$. That is approximately two back-to-back vortices, carrying each  half the energy and with  momentum scaled up by a factor $\sim c_s/c_T$ with respect to the initial state phonon. We find the amplitude
\begin{equation}
i\mathcal{M}_{L\rightarrow TT}=\frac{  c_s ^2\, p^3} {4\sqrt{w_0}} \left[ \hat \epsilon_1 \cdot \hat \epsilon_2(1+2\cos\theta^2)+\hat p\cdot \hat \epsilon_1 \hat p\cdot \hat \epsilon_2 \right ]
\label{LTT}
\end{equation}
where $\hat \epsilon_i$ represent the polarizations of the two vortices and $\cos\theta=\hat p\cdot \hat k_1$. For the decay rate we have thus roughly
\be
\Gamma_{L\rightarrow TT}\sim \frac{p^5 c_s^3}{w_0c_T^3}
\ee
corresponding to a `quality factor' $\Gamma_{L\rightarrow TT}/\omega\sim p^4c_s^2/w_0c_T^3$. A phonon with momentum 
\be
p_*\sim (w_0 c_s)^{{1}/{4}} (c_T/c_s)^{3/4}
\ee
has a width comparable to energy. We can thus identify $p_*$ as a strong interaction scale for longitudinal phonons. Notice that this scale  is parametrically smaller than the naive estimate in eq.~(\ref{strongs}). This fact is  largely a consequence of the peculiar kinematics of  the decay $L\to TT$, where starting from an initial quantum of momentum $p$, the final state quanta have a much higher momentum scale $k\sim p c_s/c_T$. This higher scale  is naturally associated with a stronger interaction strength. The fact that starting with soft quanta one can probe much shorter distances due to the large final state momentum also suggests more care with the use of the notion of effective field theory. We will elaborate briefly on this in sect.~\ref{viscosity}. Notice also that the corresponding vortex momentum strong scale is instead $k_*\sim p_*c_s/c_T\sim (w_0 c_T)^{{1}/{4}} (c_s/c_T)^{1/2}$. The computation of $TT\to TT$ scattering in the next section will show that the vortex momentum cut off is actually $\sim  (w_0 c_T)^{{1}/{4}}$, which is parametrically smaller. Then, within the resulting smaller range of validy of the effective field theory,
the process $L\to TT$ remains weakly coupled.

The results just derived display one general property of amplitudes involving vortices:
they are accompanied by at least one power of the vortex energy. \footnote{In the case of the $L\to
LT$ decay this also corresponds to a significant suppression of the amplitude, given the parametrically suppressed value of the vortex energy in that process. In the case of $L\to TT$ the energy of $L$ and $T$ modes is instead comparable.} That property directly follows from invariance under volume preserving diffeormorphisms and can be made evident by chosing a suitable field parametrization. In a neightbourhood of $\vec \phi=\vec x$ the most general field configuration can indeed be written implicitly as 
\begin{equation}
\vec \phi(x,t)= \vec g (\vec y, t) \qquad \vec y=  \vec x+\vec \pi_L(\vec x,t)
\label{factorized}
\end{equation}
where $\vec \pi_L\equiv \vec \nabla\psi( \vec x,t)$ is a longitudinal perturbation while $\vec g(\vec y,t)$ is a volume-preserving diffeormorphism generated by ``exponentiating" a transverse vector field $\vec \pi_T(\vec y,t)$ ($\vec \nabla _y \cdot \vec \pi_T = 0$). That is
\begin{equation}
\vec g(\vec y, t)= \lim_{N \to \infty}\left [ \left ( e +\frac{\vec \pi_T}{N}\right ) \circ \cdots \circ
\left (e+\frac{\vec \pi_T}{N}\right )\right ](\vec y, t)= \vec y+\vec \pi_T(\vec y,t)+O(\pi_T^2)
\end{equation}
where $e(\vec y) = \vec y$ is the identity function and $\circ$ represent function composition. Our procedure to define a finite transformation $\vec g$ starting from the infinitesimal one $\vec y+\vec \pi_T$ is just the standard exponential map of Lie groups. In the end, eq.~(\ref{factorized}) 
is a three dimensional (in field space) family of field configurations that  for $\pi_{T,L}\to 0$ reduces to the most general one $\vec x+ \vec \pi_L + \vec \pi_T$. It follows that in a neighbourhood of the identity, eq.~(\ref{factorized}) is a faithful (one to one) parametrization, acceptable to perform perturbation theory. Now a time independent $\vec \pi_T( \vec y,t)\equiv \vec \pi_T( \vec y,0)$ is just a symmetry tranformation of the action in the limit $c_T=0$. Therefore, apart from the ${\cal O}(c_T^2)$ kinetic perturbation, $\pi_T$ enters the lagrangian with at least one time derivative, and therefore amplitudes have the corresponding suppression. Notice that in the computation of $L\to LT$, where we used the simple parametrization $\vec \phi = \vec x+ \vec \pi$, that result arose via a non trivial cancellation of different terms in the amplitude.

As a further check of the above general property of amplitudes involving transverse modes,  consider the process $LL \to LT$. The relevant Feynman diagrams are
\begin{center}
\includegraphics[width=0.7\textwidth]{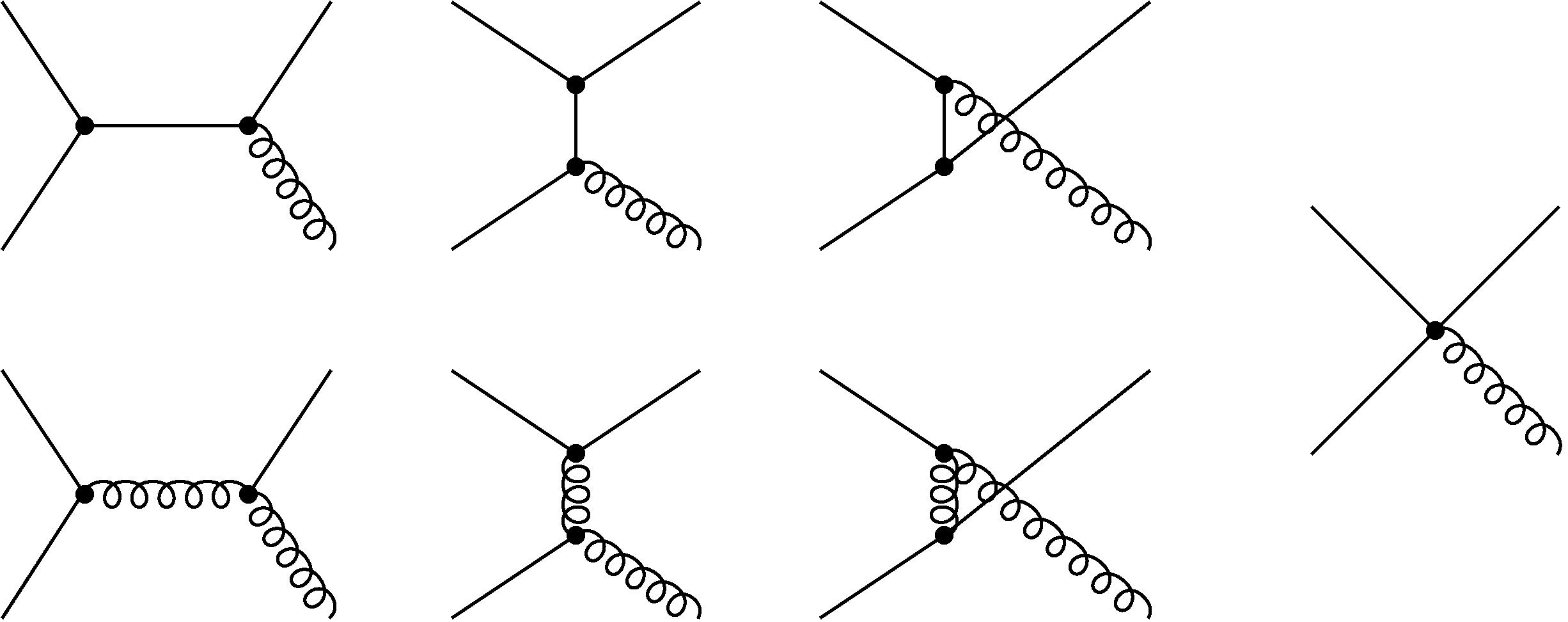}
\end{center}
The amplitude vanishes if the vortex is polarized orthogonally to the scattering plane, whereas for parallel polarization it is
\be
i{\cal M}_{LL \to LT_\parallel} = -\frac{4 i p^4 c_s c_T}{w_0} \cot \theta \big[3 -f_3/c_s^2 -2 c_s^2 - f_3 +3 c_s^4  - 2 (3- c_s^2) \cos^2 \theta    \big] + {\cal O}(c_T^2) \; ,
\ee
again vanishing, as expected, as $c_T\to 0$ with $p$ fixed. This result
corresponds to a cross-section scaling as $c_T \,$,
\be
\sigma_{LL \to LT} \sim  \frac{c_T}{c_s}\frac{1}{p^2}\left(\frac{p^4}{w_0c_s}\right)^2 \; .
\ee

The results just derived deserve one additional comment. For a classical inviscid fluid, we know that if we start with zero vorticity everywhere in space, we cannot produce any. However this is not because of some global charge conservation, but, more prosaically,
because the source of vorticity is proportional to vorticity itself. \footnote{In fact, associated with the invariance under volume-preserving diffs there are infinitely many conserved local currents and global charges \cite{DGNR}. Eq.~(\ref{Omegadot}) is a consequence of these infinitely many conservation laws, but it does not take the form of a simple conservation law for vorticity itself.} For a non-relativistic fluid:
\be \label{Omegadot}
\dot {\vec \Omega }= - \big( \vec v \cdot \vec \nabla \big) \vec \Omega+
\big( \vec \Omega \cdot \vec \nabla \big) \vec v - \vec \Omega \big( \vec \nabla \cdot \vec v \big) \; .
\ee
This is similar to, say, a scalar field theory with two fields, $\phi$ and $\chi$, with mutual interactions of the form $\phi^2 \chi^2$. In the $\chi$ e.o.m., the source term is proportional to $\chi$ itself,
\be \label{vorticity}
\frac{\delta {\cal L}}{\delta \chi} \supset \phi^2 \chi \; ,
\ee
which means that $\chi = 0$ is a perfectly good classical solution no matter what $\phi$ does. On the other hand, we know that this fact does not survive quantum mechanically. Two $\phi$ quanta in the initial state can annihilate in a $\phi^2 \chi^2$ vertex to yield two $\chi$ quanta in the final state. 
However this way one will never produce a {\em single} $\chi$ quantum if this is not already present in the initial state.
Coming back to our fluid, it is suggestive to interpret our results above in light of this analogy.
We expect that the classical non-generation of vorticity will not survive at the quantum-mechanical level.  Vortex excitations will be generically produced in scattering and decay processes, even if there are no vortices in the initial state. However, the production of a {\em single} vortex quantum should be prohibited, being immune from the aforementioned `$\phi^2 \chi^2$ effect'. Indeed consistently with this expectation, at fixed phonon momentum $p$, we found the single vortex amplitude
Eq.~(\ref{LLT}) vanishes for $c_T \to 0$, while the two  vortex  amplitude Eq.~(\ref{LTT}) does not.



\subsection{Longitudinal and transverse scattering}

Another interaction we can consider is the scattering of a longitudinal excitation and a transverse excitation. The tree level diagrams are given schematically by:
\begin{center}
\includegraphics[width=0.7\textwidth]{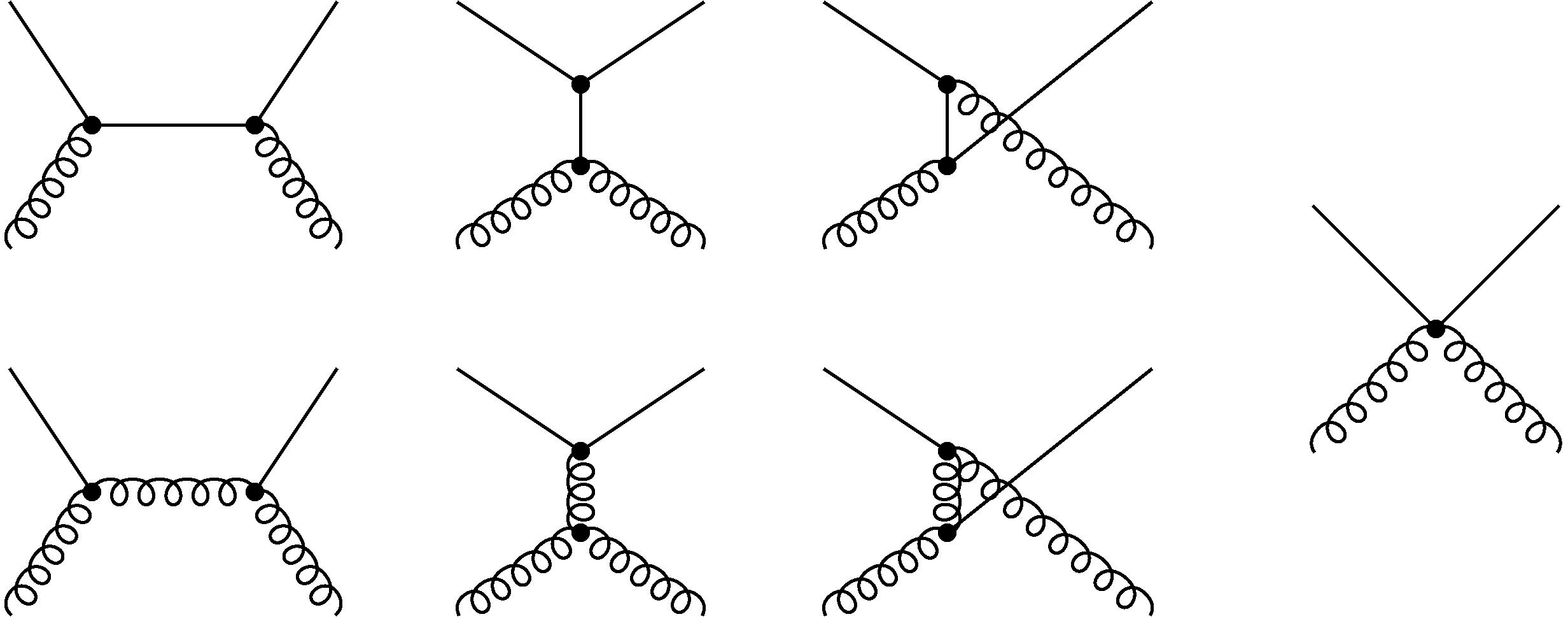}
\end{center}
Besides the usual scattering angle, there are also the additional degrees of freedom associated with the polarization of the transverse modes. The amplitude is 
\begin{equation}
i \mathcal{M}_{LT\rightarrow LT }=-\frac{i 2 c_s c_T p^4 }{w_0} \big[(\hat p \cdot \hat p') -c_s^2 \big](\hat \epsilon_1 \cdot \hat p')(\hat \epsilon_2 \cdot \hat p)+\mathcal{O}(c_T^2)
\end{equation}
Note that $\mathcal{M}\propto c_T$, consistent with the general argument presented in the previous section. It should also be noted that there is explicitly no  dependence on $f_4$ and  $f_3$. This result can be easily understood using the field parametrization discussed in the previous section. The terms proportional to $f_4$ and $f_3$ only depend on the longitudinal field $\pi_L$. Then $f_4$ obviously does not contribute to $LT\to LT$ while $f_3$ can only contribute via the second diagram in the upper line in the figure, which is clearly of order $c_T^2$.
 Again, in the standard parametrization all these results follow from a non-trivial set of cancellation, which also represent a check of our computations.
 
 The infinitesimal cross-section is given by 
\begin{equation}
d\sigma=\frac{1}{2}\sum_{\text{initial $\epsilon$}} \sum_{\text{final $\epsilon$}}\left( \frac{1}{p^2}\right)\left( \frac{p^4}{4\pi w_0 (c_T+c_L)}\right)^2 \big\{ \big[(\hat p \cdot \hat p') -c_s^2 \big](\hat \epsilon_1 \cdot \hat p')(\hat \epsilon_2 \cdot \hat p)\big\}^2d\Omega
\end{equation}
Here we are averaging over the incoming polarizations (hence the $1/2$) and summing over the final ones. A good basis to do this in would be parallel to the scattering plane and perpendicular to the scattering plane. As we can see from the form of $\mathcal{M}$, any perpendicular component of the polarization does not contribute to the amplitude. Putting everything together, keeping all powers of $c_s$, we have the total cross section:
\begin{equation}
\sigma=\frac{1}{105\pi}\frac{1}{p^2} \left( \frac{p^4}{ w_0 c_s}\right)^2 [1+7c_s^4] +\mathcal{O}(c_T)
\end{equation}
We now see the importance of $\mathcal{M}_{LT\rightarrow LT} \propto c_T$. This dependence on the ``transverse speed of sound" is necessary to avoid a divergent physical quantity (here the scattering cross-section) as $c_T\rightarrow 0$. Remarkably, the strong-coupling scale for this process is the same as for purely longitudinal scattering---cf.~eq.~(\ref{sigmaL}). Note also the independence of the cross section on $f_3$ and $f_4$. This cross section is generic for all fluid types, regardless of the particular functional form of $f(\sqrt{B})$. Its dependence on the particular fluid model only comes through the speed of sound.


\subsection{Transverse $2 \to 2$ scattering}

 As we can see from formula (\ref{sigma}) for the differential scattering cross section, the more transverse incoming and outgoing states the more a rate  could possibly diverge as $c_T\to 0$. The results so far agree with that expectation. The only problematic quantity we encoutered is the rate for $L\to TT$, which has however a peculiar (singular) kinematics as $c_T\to 0$. The processes with smooth kinematics were instead found to have a well behaved rate. In the $T+L \rightarrow T+L$ cross section we picked up a $c_T^{-1}$ from the phase space of the outgoing excitation and we picked up a $c_T^{-1}$ from one of the $1/2E$ normalization factors. It was thus critical that  $|\mathcal{M}_{LT\rightarrow LT}| \propto c_T^2$ in order that the cross section be well defined.

From eq.~(\ref{sigma}) we can see that for transverse 2 to 2 scattering $\frac{d\sigma}{d\Omega}\propto c_T^{-6}|\mathcal{M}|^2$. In order that our physical process be finite in the $c_T\rightarrow0$ limit we need (at least) $\mathcal{M} \propto c_T^3$. We find that this is not the case. In fact, simple power counting using the parametrization discussed in section 4.2 indicates  the amplitude will be suppressed (at least) as $c_T^2$. This expectation is confirmed working  in the standard  parametrization: by dramatic cancelations among Feynman diagrams the zeroth and first order terms vanish, and the leading non-trivial term is of order $c_T^2$. The necessary tree-level Feynman diagrams are:
\begin{center}
\includegraphics[width=0.7\textwidth]{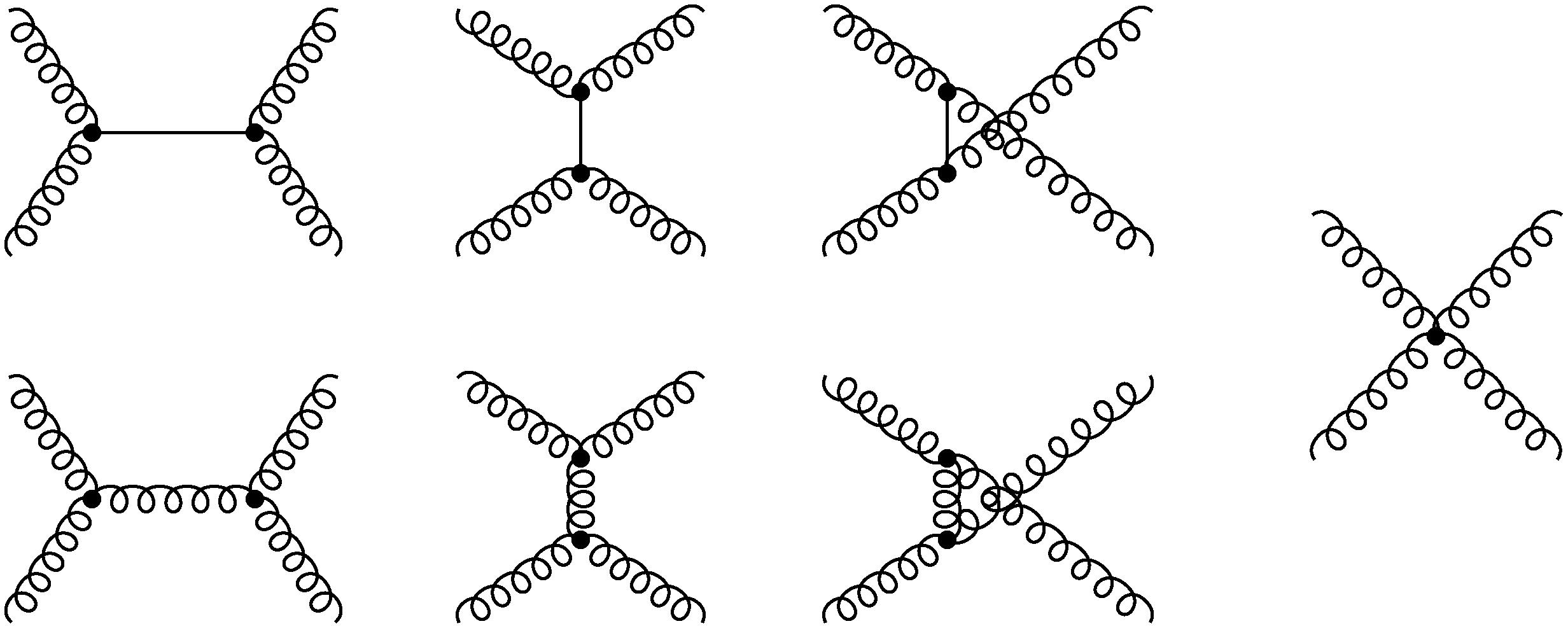}
\end{center}
In the CM frame with $\parallel$ and $\perp$ denoting polarizations parallel and perpendicular to the scattering plane, respectively,  at lowest order in $c_T$ the amplitude is
\be
i \mathcal{M}_{TT\rightarrow TT} =\frac{i k^4 c_T ^2}{w_0} \times
\left\{ \begin{array}{ll}
\cos 2\theta & \mbox{ for } \parallel \parallel \to \perp \perp \; , 
\perp \perp \to \parallel \parallel \\
\sfrac12 \big( \cos \theta - \cos 2 \theta \big) & \mbox{ for } \parallel \perp \to \parallel \perp \\
\end{array}
\right.
\ee
and zero for all other combinations of polarizations. In the $\parallel \perp \to \parallel \perp $ case, $\theta$ is the angle between the two $\parallel$-polarized phonons.
%
%
Note that, as in the previous physical process, there is no dependence on $f_3$ and $f_4$. So, once again, this process is generic for all fluid models regardless of the details of $f(\sqrt{B})$.

After squaring the amplitude, we average over incoming polarizations and sum over the final ones. 
We have:
\begin{equation}
\frac{1}{4}\sum_{\text{initial $\epsilon$}} \sum_{\text{final $\epsilon$}}|\mathcal{M}|^2=\frac{1}{4}\left( \frac{k^4 c_T^2}{w_0} \right)^2 \big[2+\sfrac{1}{2} \cos 2\theta +\sfrac{3}{2} \cos4\theta \big] \; ,
\end{equation}
and so the total cross section, including a $1/2$ symmetry factor, is 
\begin{equation}
\sigma_{TT\rightarrow TT}=\frac{1}{256 \pi} \left( \frac{13}{15}\right)\frac{1}{k^2}\left( \frac{k^4}{w_0 c_T} \right)^2 \; .
\end{equation}
As we can see, it blows up as we take $c_T\rightarrow 0$ indicating that our theory is breaking down. We emphasize once again the absence of the free parameters $f_3$ and $f_4$ in this result, which implies that the pathology just unveiled cannot be avoided by a judicious choice of their values.

\section{The infrared situation}\label{IR}

Our $S$-matrix analysis indicates that the transverse degrees of freedom are strongly-coupled
at arbitrarily low energies. However the  strong-coupling phenomenon we unveiled is quite peculiar. We deformed the theory in the far IR by introducing a small deformation parameter $c_T$. This changes the asymptotic states of the theory, and we discovered that some of these get strongly coupled in the UV, at an energy scale that drops to zero when we recover the original theory---the limit $c_T \to 0$. A vanishing {\em ultraviolet} strong-coupling energy scale suggests that our problem is probably more properly thought of as an {\em infrared}
one---we may be approaching strong-coupling from the wrong side! That is, in the deformed theory at finite $c_T$ we encounter strong coupling in moving to high energies, but since the deformed theory differs from the original one at low energies, it may be that the strong coupling scale is in fact a divide between the two theories---that there is no regime where the two theories look alike. If we stick to the original description, without ever introducing $c_T$, we may realize that we have some form of strong coupling in the IR.

The distinction we are putting forward may sound like a matter of definition, but it is not. A theory that becomes strongly coupled in the UV is simply not defined at energies of order of the strong-coupling scale and above---it needs infinitely many parameters for its definition. If this were the case for us, our theory would not be consistent, in any energy range.
On the other hand, there are a number of  ways in which perturbation theory can break down in the IR without impairing the consistency of a theory. There is for instance real QCD-like strong coupling, where perturbation theory does break down but the non-perturbative theory is perfectly well defined---it is just hard to solve! Or there are QED-like infrared divergences, which can be tamed by focusing on suitable infrared-safe observables, for which the perturbative expansion applies. Or there may be huge quantum IR fluctuations without necessarily implying large interactions, like for instance for would-be Goldstone bosons in $1+1$ dimensions \cite{coleman}. This would signal a bad identification of the theory's vacuum state. 
And in general, to ascertain the consistency of an IR-problematic theory, at least in certain energy and momentum regimes, one can just put the system in a finite-size `box' and consider its time evolution for a short time, in which case perturbation theory typically does not even break down.

The first signal that our problem with the transverse degrees of freedom may be infrared in nature comes from considering quantum fluctuations about the semiclassical vacuum state with $\langle \phi^I \rangle = x^I$.
An order parameter that conveniently quantifies the amount of spontaneous symmetry breaking in a manifestly  translationally invariant fashion is
\be \label{order}
\langle \di_\mu \phi^I \rangle = \delta_\mu^I \sim 1 \; .
\ee
It is straightforward to estimate quantum fluctuations in this quantity. We decompose the fields as $\phi^I = x^I+ \pi^I$, and from the $\pi^I$ propagators,
\be
\langle T \pi^I(x) \pi^J(y) \rangle \to \frac{1}{w_0} \cdot \frac{i P_L^{IJ}}{\omega^2 - c_L^2 p^2 + i \epsilon} + \frac{1}{w_0} \cdot \frac{i P_T^{IJ}}{\omega^2 +i\epsilon} \; ,
\ee
where $P_L$ and $P_T$ are the longitudinal and transverse projectors, we get
\bea
\langle \di_i \pi^I \, \di_j \pi^J \rangle & \sim & \frac{1}{w_0} \frac{p^5}{\omega} \label{ij} \\
\langle \di_0 \pi^I \, \di_j \pi^J \rangle & \sim &  \frac{1}{w_0}  p^4 \label{0i} \\
\langle \di_0 \pi^I \, \di_0 \pi^J \rangle & \sim &  \frac{1}{w_0}  p^3 \omega \label{00} 
\eea
These are real-space correlators, and in the right-hand sides the dimensionless, order-one part of the Fourier transform, $\int \! d \Omega \, d \log p \, d \log \omega \, e^{i ( \dots)}$, is understood. Also we are considering considerably off-shell $(\omega, p)$ pairs, by taking for instance the separation in real space to be space-like with the respect to the sound speed (i.e., by working in Euclidean space.)
The correlators (\ref{0i}, \ref{00}) behave in essentially the same way as for standard field theories in dimensions higher than $1+1$: quantum fluctuations in order parameters are damped at low momenta and low energies, and as a consequence in the IR  there is spontaneous symmetry breaking. However the correlator (\ref{ij}) ruins this familiar picture: for fixed momentum, it is IR-divergent. Equivalently, in the $p$-$\omega$ plane there is a sector extending all the way to $p=0$, $\omega=0$ where quantum fluctuations in our order parameter (\ref{order}) are huge. This reminds us of Coleman's theorem \cite{coleman} in $1+1$ dimensions, and suggests that  our strong-coupling problems may stem from an incorrect definition of the vacuum. In other words, we have been assuming all along that doing perturbation theory about the semiclassical vacuum $\phi^I = x^I$ is sensible, but eq.~(\ref{ij}) is somehow telling us that quantum fluctuations want to dismantle this state. As an aside, notice  that we have all reasons to believe that Lorentz boosts {\em are} spontaneously broken: since the correlators (\ref{0i}, \ref{00}) are damped in the IR,
it seems that only the breaking of the spatial symmetries (translations, rotations, volume-preserving diffs) is affected by our phenomenon. More concretely, we can compute quantum fluctuations in an order parameter that is invariant under all symmetries but Lorentz boosts, like for instance the fluid velocity $u^\mu$. Classically in the ground state we have $u^\mu = (1,\vec 0)$. At first order in the $\pi$ field the fluctuation reads $\delta u^\mu = (0, -\dot {\vec \pi})$. From eq.~(\ref{00}) we thus get the velocity-velocity correlator,
\be
\langle \delta u^\mu \, \delta u^\nu \rangle \sim \delta^\mu_i \delta^\nu_j  \, \frac{1}{w_0}  p^3 \omega \; ,
\ee
which is damped in the IR, thus signaling that quantum fluctuations do {\em not} want to restore Lorentz invariance.

Several natural questions arise: How do we check the above statements more concretely? Can we identify the correct vacuum state? Does it support a well defined perturbative theory? And perhaps the most physically relevant: is there a semiclassical limit where we recover classical hydrodynamics?
To start addressing these questions, we step back from our fluid case and consider a much simpler system with somewhat similar features: the free particle in quantum mechanics.

\subsection{The free quantum particle}\label{QMparticle}

Consider a free particle living on a line. Classically, $x = x_i = {\rm const}$ is a perfectly good state. However, quantum mechanically we know that if we start from a state localized around $x_i$, time-evolution will make the wave-function spread out, and at very late times the state will be totally delocalized. Related to this, the ground state of the theory wants to have a constant wave-function throughout the whole line. (This is non-normalizable for an infinite line, but for simplicity we can replace the line by a very large circle.) The spontaneous breaking of translations that we see classically, $x = {\rm const}$, quantum mechanically is gone. 

Given the system's simplicity, there are many equivalent ways to describe this phenomenon  quantitatively. One that will prove readily exportable to the fluid case, is the path integral one. Consider, in the path-integral representation, the amplitude for evolving from $x_i$ at $t=0$ to $x_f$ at $t=T$:
\be
\langle x_f , T \, | x_i, 0 \rangle = \int_{x(0) = x_i}^ {x(T) = x_f} {\cal D} x \, e^{iS[x]} \propto \exp{\frac{i \,(x_f-x_i)^2}{2T}} \; . \label{free}
\ee
We are using $\hbar=1$ units. Also for simplicity we are setting the particle's mass to one. The exponent is of course the classical action for the solution interpolating between $x_i$ and $x_f$ in time $T$. The prefactor that we omitted is the determinant of the kinetic operator about the classical solution. Since we have a free theory such an object does not depend on the solution under consideration, or on the overall distance $(x_f-x_i)$, and will play no role in our discussion---it is just an overall normalization factor. 

We see that, for a given $T$, at small distances $|x_f-x_i| \lesssim \sqrt T$ the amplitude is essentially constant, whereas it starts oscillating rapidly when we move to much larger distances, $|x_f-x_i| \gg \sqrt T$. We are tempted to conclude from this that the width of the wave-function grows like $\sqrt{T}$, but this is premature because the amplitude we got is a pure phase at all distances, which means that the probability is uniform in $x_f$. In other words, as soon as $T$ is bigger than zero, the wave function is completely delocalized. This is clearly not what we expect physically, and is an artifact of having chosen an initial $\delta$-like wave-function: a $\delta(x)$ has infinite momentum spread, which means that  in zero time the particle is going to be everywhere with non-vanishing probability.
If we instead choose a more reasonable state, say a gaussian centered at $x_i=0$ with width $\sigma$, the final wave function we get is:
\be
\psi(x_f, T) = \int \! dx_i \, \psi(x_i, 0) \,  \langle x_f , T \, | x_i, 0 \rangle \propto 
\left\{ \begin{array}{lr}
\exp{-\frac{x_f^2}{2 \sigma^2}} \; ,  & T \ll \sigma^2 \\
\\
\exp{-\frac{x_f^2 \, \sigma^2}{T^2}} \; ,  &  T \gg \sigma^2 
\end{array} \right.
\ee
where we ignored overall normalization constants, as well as factors that do depend on $x_f$ but are pure phases.
We thus see that at early times the wave-function width is---not surprisingly---dominated by the initial width $\sigma$, whereas at late times it grows linearly with $T$. 

We can turn the problem around and ask: suppose we want to have a somewhat localized state that looks stationary over a time period of order $T$. What is the minimum wave-function width we should allow for? Answer: $\sqrt{T}$. As a consequence, in the long-time limit there cannot be spontaneous symmetry breaking.
With hindsight, we can go back to our original amplitude eq.~(\ref{free}) and learn how to read it properly.
Despite being purely imaginary, the exponent really tells us the minimum uncertainty we should allow in the initial position in order for our state to look approximately stationary over a time of order $T$.

Before moving on to the field theory case, it is worth pointing out that if we have $N$ free particles we can run the above computation independently for each of them, since all amplitudes factorize. This means that, roughly speaking, each particle can afford an ${\cal O}(1)$ action, so that the whole system can explore trajectories with an ${\cal O}(N)$ action.


\subsection{Coleman's theorem from the path integral}\label{coleman_section}

We can now apply the same logic to a massless free field theory in $d+1$ dimensions, 
\be
S[\phi] = \int \! d^{d+1} x \, \sfrac12 (\di \phi)^2 \; ,
\ee
and recover Coleman's theorem in the path-integral language. We  have a shift symmetry $\phi \to \phi + c$, which classically is spontaneously broken by any Poincar\'e invariant configuration $\phi (x) = {\rm const} $. Without loss of generality we can set the constant to zero, and ask whether quantum fluctuations tend to disrupt this configuration. Coleman teaches us that this will be the case for $d=1$.

We want to compute the path integral
\be
\langle f (\vec x), T \, | 0, 0 \rangle = \int_{\phi(0) = 0}^ {\phi(T) = f(\vec x)} {\cal D} \phi \, e^{iS[\phi]} \propto e^{iS[\phi_{\rm cl}]}  \; , \label{coleman}
\ee
for a generic final configuration $f(\vec x)$. Given what we learned for the free particle on a line, the result will tell us which are the field configurations that inevitably get populated after a time $T$ if we start at $t=0$ with a wave-functional for $\phi(\vec x)$ centered around $\phi=0$. To compute $S[\phi_{\rm cl}]$, we have to solve the classical equations of motion with the given boundary conditions. In (spacial) Fourier space the immediate solution is:
\be
\hat \phi_{\rm cl}(\vec k, t) = \hat f (\vec k) \cdot \frac{\sin kt}{\sin kT} \; ,
\ee
where $\hat f (\vec k)$ is the final configuration's Fourier transform. The action is
\be
S[\phi_{\rm cl}] = \int \! d^d  k \,  \big| \hat f (\vec k) \big|^2  \cdot k  \cot kT \; .
\ee
We can for simplicity consider the two regions $k \ll 1/T$ and $k \gg 1/T$ separately, and ignore what happens at intermediate momenta. As to the former region, we can set $\cot kT \to 1/kT$, so that we have
\be
S(k \ll 1/T) \simeq \frac{1}{T} \int^{1/T} \! d^d  k \,  \big| \hat f (\vec k) \big|^2  \; .
\ee
The high momentum region, however, contains all the poles of the cotangent. These poles are just enforcing the right periodicity for the corresponding momenta. That is, for a given $T$ all modes with $k = n \pi/T$ should go back to zero after a time $T$, as enforced by the perfect harmonicity of our system. As a consequence, if  the final $\hat f (\vec k)$ does not vanish for these special momenta, the corresponding action is infinite. To have a finite action $\hat f (\vec k)$ has to have infinitely many zeroes, at the right locations. Apart from this peculiarity,  when we integrate over momenta much larger than $1/T$ the cotangent behaves like a number of order one, and to get an estimate for the action we can just ignore it. We thus have
\be
S(k \gg 1/T) \sim \int_{1/T} \! d^d  k \,  \big| \hat f (\vec k) \big|^2 \cdot k  \; .
\ee

We can now consider a specific final configuration $f(\vec x)$. We choose it to be localized in a region of size $L$, with Fourier momenta of order $1/L$ and with typical magnitude $\bar f$. As to the final remark of sect.~\ref{QMparticle}, here we are considering just a few independent degrees of freedom---the Fourier modes with $k \sim 1/L$ in a volume of size $L$---so that they get spontaneously excited to the desired level $\bar f$ only if the corresponding action in (\ref{coleman}) is of order one, or smaller.
At early and late times we get, respectively
\be \label{LandT}
S(T \ll L) \sim \frac{L^d}{T} \bar f^2 \; , \qquad S(T \gg L) \sim L^{d-1} \bar f^2
\ee
Our path-integral formula (\ref{coleman}) at late times thus yields 
\be
\langle f (\vec x), T \, | 0, 0 \rangle  \sim \exp {i \, L^{d-1} \bar f \,^2} \; , \qquad T \gg L
\ee
This is Coleman's theorem in our language. For $d > 1$,  the amplitude for finding the system significantly away from $\phi = 0$ (large $\bar f$) in  bigger and bigger regions becomes smaller and smaller. Put another way, the typical spread of the field decreases with distance scale. This is the standard behavior of a field theory in high dimensionality. On the other hand, for $d=1$ the $L$-dependence drops out, and we are left with a finite amplitude of finding an order-one $\bar f$ on all scales. In fact we know that a more careful estimate shows that the typical spread of the field {\em grows} logarithmically with distance scale---as usual one cannot get the logs right by performing  Fourier transforms by dimensional analysis.
Our analysis for the fluid will be insensitive to such minuti\ae.

\subsection{Back to the fluid}

For our fluid, it is easy to convince oneself that the situation is more similar to the free particle case than to a $1+1$ standard field theory. The reason is simply that all the peculiarities associated with the transverse excitations in a perfect fluid stem from their not having a gradient energy, that is, from their behaving like infinitely many free particles. Of course, this is a linearized statement. At the non-linear level, for large vorticous excitations, interactions become important and the free-particle approximation breaks down. But, in fact, it is easy to see that at low fluid velocities we can really describe the dynamics of vortices as a system of free particles constrained to move on a quite non-trivial infinite-dimensional manifold---the group of volume preserving diffeomorphisms.

To see this,  we have to take the incompressible fluid limit. Of course, a fluid's compressibility is not a dimensionless quantity. It is a measure of the pressure gradient needed to sustain a given density gradient.
Since $dp / d \rho$ is the (squared) speed of sound, incompressibility is not an intrinsic property of the fluid, but rather emerges in the appropriate kinematical regime: {\em any} fluid behaves as an incompressible one at low enough fluid velocities. In this limit, one can consistently restrict to the vortex sector of the theory, i.e.~to configurations $\phi(\vec x, t)$ that at fixed time are volume-preserving diffs of $\vec x$, as we now show. 

\subsubsection{Integrating out sound}
It is easy to see that for low speeds and accelerations (i.e.~weak time-dependence of $\phi$), integrating out classically the compressional modes introduces new Lagrangian terms for the vortices that are of order $\di_t^4$ and higher, whereas the vortex dynamics that we get 
by simply ignoring the compressional modes starts at order $\di_t^2$. 
Indeed, consider the action for our fluid at small velocities. It is more convenient to use the $\vec x(\vec \phi, t)$ parameterization for the fluid, i.e.~to work in comoving coordinates. Moreover, it is instructive to reinstate the speed of light $c$ explicitly, to disentangle more transparently relativistic effects from those we are after. We have 
\bea
S & =  & - w_0 c^2 \int \! d^3 \phi dt \, \det J \,f\big( ( \det J^{-1}) \sqrt{1- v ^2/c^2} \big) \\
& \simeq & w_0 \int \! d^3 \phi dt \, \big[ -c^2 \det J\, f( \det J^{-1})\\ 
 & &+  f'( \det J^{-1}) \, [ \sfrac12 v ^2 \, + \sfrac18 v^4/c^2 ] \, - \sfrac18  \det J^{-1} f''( \det J^{-1}) \, v ^4/c^2 \,  \big]  \label{incompressible}\; ,
\eea
where $J^i {}_j$ is the Jacobian matrix $\di x^i / \di \phi^j$, and $\vec v = \di_t {\vec x} (\vec \phi, t)$ is the fluid's velocity field. We are giving $w_0$ units of a mass density, and $f$ is a dimensionless function of its dimensionless argument. The above expression for the action is derived in the Appendix (eq.~(\ref{expandL})), or in ref.~\cite{DGNR}. Now, for nearly incompressible flows, we parameterize the fluid's configuration as
\be
{\vec x} (\vec \phi, t) = {\vec x}_0 (\vec \phi, t) + \delta \vec x (\vec \phi, t) \;, 
\ee
where $\vec x_0$ is, at fixed $t$, a volume-preserving diff of $\vec \phi$,
\be
(J_0)^i {}_j \equiv \frac{\di x_0^i}{\di \phi^j} \; , \qquad \det J_0 = 1 \; ,
\ee
and $\delta \vec x$ vanishes in the limit of negligible time-dependence of $\vec x_0$. The vorticose motion and the fluid's compression are parameterized, respectively, by $\vec x_0$ and $\delta \vec x$. At lowest order in $\delta \vec x$, this is achieved by imposing that $\delta \vec x$ be longitudinal as a function of $\vec x_0$, that is
\be
\delta \vec x = \vec \nabla_0 \psi(\vec x_0, t) \; , \qquad 
\vec \nabla_0 \equiv \frac{\di}{\di \vec x_0} = (J_0^T)^{-1} \cdot \frac{\di}{\di \vec \phi} \; .
\ee 
This corresponds to decomposing the fluid configuration into transverse and longitudinal displacements, at non-linear order in the transverse ones but at linear order in the longitudinal ones.
Of course, we may always add background sound waves to our fluid, in which case $\delta \vec x$ is not necessarily small; but here we are after the compressional modes necessarily associated with a given system of vortices by the dynamics. In terms of the above decomposition we have
\be
\det J = 1 + \nabla_0^2 \psi + \sfrac12 \big[ \big( \nabla_0 ^2 \psi \big)^2 -  \big(\nabla_0 ^{i} \nabla_0 ^{j} \psi \big)^2 \big]+ \dots \; ,
\ee
and for the velocity field
\be
\vec v \equiv \di_t  \, {\vec x} (\vec \phi, t) = \vec v_0 + \di_t \vec \nabla_0 \psi \; .
\ee
Notice that these partial time-derivatives are to be evaluated at fixed comoving position $\vec \phi$, i.e.~they are Lagrangian derivatives $D/Dt$. To integrate out (classically) $\delta \vec x$, we expand the action at quadratic order in it. Keeping in mind that $\psi$ will be suppressed by time derivatives, the leading terms in a small $\di_t$ expansion are
\be
S \simeq  S_{0} + w_0 \int \! d^3 x_0 dt \, \Big[ - \sfrac 1 2 c_s^2 \, ( \nabla_0^2 \psi)^2  - \sfrac12  c_s^2 \, (\nabla_0 ^2 \psi )\,  \frac{v_0 ^2}{c^2} -  \vec \nabla_0 \psi \cdot \frac{D}{Dt} \vec v_0 \Big]  \; ,
\ee
where $S_0$ is the $\delta \vec x$-independent part of the action, we used $f'(1) =1$, $f''(1) = c_s^2/c^2 $ (see sect.~\ref{eft}), and we changed coordinates from $\vec \phi$ to $\vec x_0$---for which the Jacobian determinant is one, and upon which we dropped total derivative terms. 
Notice, crucially, the absence of a $( \nabla_0 ^2  \psi)^2 $ term multiplied by $c^2$ rather than by $c_s^2$, which can be parameterically smaller than $c^2$. Naively such a term should be there, from expanding the first term in eq.~(\ref{incompressible}). Yet we know that precisely the coefficient of the compressional mode gradient energy defines the propagation speed of sound. Therefore terms of order $c^2 ( \nabla_0^2 \psi)^2$ must cancel out, and indeed they do---like they did for eq.~(\ref{free_action}). 
Moreover, the second term in the action is the only one suppressed by  $c^2$, and is clearly associated with relativistic effects. For a non-relativistic fluid with $c_s \ll c$, it can be safely ignored. On the other hand, for relativistic fluids ($c_s \sim c$) with non-relativistic flows ($v \ll c$), such a term should be kept, being of the same order as the other ones.
Finally, at zeroth order in $\delta \vec x$ we can replace the Lagrangian derivative $D/Dt$ with that associated with the incompressible flow,
\be
\frac{D_0}{D_0 t} \equiv \frac{\di}{\di t} \Big|_{{\rm const } \: \vec x_0} + \vec  v_0 \cdot \vec \nabla_0 \; .
\ee
Varying the action with respect to $\psi$ and solving for it, we get
\be
\psi = -\frac12 \frac{1}{\nabla_0^2}\frac{v_0 ^2}{c^2} +\frac{1}{c_s^2} \frac{1}{\nabla_0^4} \vec \nabla_0 \cdot \frac{D_0}{D_0 t} \vec v_0 \; .
\ee
Plugging back into the action (and dropping the subscript `$0$') we get the effective action for the vortex variables,
\bea
S_{\rm eff} & = & S_{0}  + \Delta S \\
S_{0} & = &w_0 \int \! d^3 \phi \, dt \, \Big[\frac{1}{2} v ^2 + \frac{1}{8} \frac{v^4}{c^2}- \frac{1}{8} \frac{c_s^2v^4}{c^4} \Big]  \label{free_vortices}\\ 
\Delta S & \simeq & w_0 \int \! d^3 x dt  \, \Big[ 
\frac1{8} \frac{c_s^2 v^4}{c^4} -  
\frac1{2} \frac{v^2}{c^2} \Big( \frac{1}{\nabla^2}  \vec \nabla \cdot \frac{D}{Dt} \vec v\Big) +
\frac1{2} \frac{1}{c_s^2} \Big(  \frac{1}{\nabla^2} \vec \nabla \cdot \frac{D}{Dt} \vec v\Big)^2 \Big]
\eea
where in $S_{0} $ we dropped the rest-mass, $\vec v$-independent contribution. As advertised, integrating out the compressional modes introduces ${\cal O}(\di^4_t)$ corrections to the vortex action. At small velocities and accelerations we can just restrict our original Lagrangian to the volume-preserving configurations, that is we can just  use the lowest order term in $S_{0}$.
In this approximation, and in comoving coordinates, the dynamics are free, but of course we have a very non-trivial constraint on the configuration space: $\vec x(\vec \phi, t)$ cannot leave the space of volume preserving diffs.

Before moving on, it is worth pointing out a couple of features of the computation we just performed. The ${\cal O}(\di_t^4)$ correction we computed is local in time but non-local in space. Some form of non-locality was to be expected since we are integrating out a massless degree of freedom. Yet locality in time follows from our expanding in powers of $\di_t$. That is, our procedure is equivalent to expanding the sound-wave propagator at low frequencies but finite momenta:
\be
\frac{1}{\omega^2- c_s^2 k^2} \to -\frac{1}{c_s^2 k^2} + \frac{\omega^2}{c_s^4 k^4} + \dots \; , \qquad \omega \ll c_s k \; .
\ee
At any finite order such an expansion is analytic in $\omega$, and thus local in time. In this sense the momentum here is playing the role usually played by the mass when one integrates out a heavy particle in a relativistic QFT. For a given typical frequency $\omega$, at momenta of order of $c_s \omega$, that is at distances of order $1/c_s \omega$, we expect the expansion to break down. Indeed, we know that no matter how slow our vortices, since sound waves are gapless, there will be some sound emission. To see (actually to hear) this sound one has to go to the so-called wave-zone, which is indeed $1/c_s \omega$ away from the sources. In a finite volume however, the wave-zone may just not be there. Equivalently, there is a gap in the frequency of sound-waves. Moreover, even in infinite volume, since interactions involve derivatives, the sound emission will be suppressed at very small frequencies and momenta. One can therefore make sound emission arbitrarily slow by working at low vortex speeds, thus making the procedure of integrating out sound-modes perfectly sensible. It would be interesting to carry out this program more systematically, and characterize sound wave-mediated interactions between vortices and the sound emission by vorticose motions.
We leave this for future work.

\subsubsection{The absence of spontaneous symmetry breaking}

We can now formulate precisely the question of spontaneous symmetry breaking for our fluid. We want to compute the amplitude to propagate from $\vec x = \vec \phi$ to $\vec x = \vec \xi(\phi)$, with $\vec \xi$ a volume preserving diff:
\be
\big \langle \vec \xi (\phi  ), T \, \big | \vec \phi, 0 \big \rangle = \int_{\vec x(0) = \vec \phi}^ {\vec  x(T) = \vec \xi( \phi)} {\cal D} \vec x \, e^{iS[\vec x]} \propto e^{iS[\vec x _{\rm cl}]}
\ee 
If at late times all volume-preserving diffs get populated with equal probability, there is no spontaneous symmetry breaking and our perturbative analysis in the broken phase is not applicable.

Unfortunately, we are not able to carry out this computation for a completely generic final configuration $\vec \xi (\phi)$: given the simple free dynamics of the system, we can confidently say that there will be  a  classical solution with time-independent velocity field $\vec v(\vec x)$ evolving from $\vec x = \vec \phi$ to $\vec x = \vec \xi(\phi)$. However, because of the complexity of the manifold where motion takes place, determining the velocity field that connects the initial and final configuration would be quite hard. And without knowledge of the velocity field, we cannot compute the classical action. 

Nevertheless, to assess the question of spontaneous symmetry breaking we do not need to be completely general. We can focus, for instance, on sample vortex configurations with a high degree of symmetry. For example, we can consider $\vec \xi(\phi)$ to be a rotation around the $z$-axis, of an angle $\Delta\varphi$ that depends on the distance $r$ from the axis (dropping to zero above some distance $R$), localized in a region of length $L \gg R$ in the $z$ direction.
\footnote{Similarly to the sample $f(\vec x)$ considered in sect.~\ref{coleman_section}, the final configuration considered here corresponds to exciting an ${\cal O}(1)$ number of degrees of freedom. In a region of sizes $L$, $R$, and  $R$ we are considering essentially just the `fundamental harmonic', with $k_z \sim 1/L$ and $k_x \sim k_y \sim 1/R$. This makes the comment at the end of sect.~\ref{QMparticle} immaterial for our purposes.}
The classical solution connecting the initial configuration to this is a vortex in constant rotation with $r$-dependent angular velocity,
\be
\omega(r) = \frac{\Delta \varphi(r)}{T}
\ee
localized in a cylinder of radius $R$ and height $L$. From eq.~(\ref{free_vortices}), the classical action for this solution is
\be
S_{\rm cl} [\Delta \varphi] \simeq 2 \pi \int \! dz r d r dt \, \sfrac 12 w_0 r^2 \omega^2 (r) = \frac{w_0}{T} \pi  \int \! dz d r \,  r^3 \Delta \varphi^2 (r) \sim w_0 \frac{R^4 L \bar{\Delta \varphi}^2}{T} \; , 
\ee
where $\bar{\Delta \varphi}$ is the typical overall rotation of the final configuration in the region of interest. The amplitude we are interested in therefore is
\be
\big \langle \mbox{rotation of } \bar{\Delta \varphi} \mbox{ in } R,L \, ; T \, \big | \vec \phi, 0 \big \rangle \sim \exp i \frac{w_0 R^4 L }{T} \bar{\Delta \varphi}^2 \; .
\ee
Angles of order one get populated in a cylindrical region of any given size $R$, and $L $ if we wait long enough:
\be \label{T}
T \gtrsim w_0 R^4 L \; . 
\ee
At late times the symmetry is completely restored.
Notice how the estimate (\ref{T}) matches precisely what we could have inferred from the linearized statement~(\ref{ij}).

\section{Quantum viscosity?}\label{viscosity}

Even if the above arguments alleviate the concerns raised by the perturbative analysis of sect.~\ref{processes}, we are still left with indications that the effective theory at hand may not be unitary.

A standard effective field theory can be unitary at low energies thanks to the decoupling of the short distance degrees of freedom. That is, thanks to the fact that to excite the microscopic degrees of freedom one is neglecting/being agnostic about, one needs a non-zero energy. There is a gap, and as long as one works below the gap, the long-distance degrees of freedom are sufficient to parameterize the dynamics.

In our case, we are not in such a good shape. To excite very microscopic vorticose deformations of the fluid we need no energy at all. High momenta are not associated with high energies. This is also evident from eq.~(\ref{ij}), where energy and momentum play strikingly different roles. 
This suggests that for any given cutoff in momentum space, the effective theory may not be exactly unitary. No matter how small the energy of the process under consideration, there may be a non-trivial probability flow across the cutoff. From the effective theory viewpoint this should look like dissipation. In fact, for classical turbulence in a viscous fluid this is exactly what happens---viscosity drives vorticity from large scales to smaller and smaller ones, down to the UV cutoff of the fluid description (the mean free path of the underlying microscopic system.) Could it be that in our case we have some sort of quantum contribution to viscosity due to this non-decoupling of micro-vortices? How can we test this conjecture?

\section{Discussion and Outlook}

Our findings raise more questions than they answer. The perturbative analysis about the naive, semiclassical vacuum $\phi^I = x^I$ indicates that the ordinary fluid effective field theory is strongly coupled at all scales, and thus inconsistent (sect.~\ref{processes}). On the other hand, a more careful non-perturbative study of the theory's quantum-mechanical vacuum shows that this has essentially nothing to do with the semiclassical one, and suggests that the naive perturbative degrees of freedom and their dynamics have no quantum-mechanical counterpart (sect.~\ref{IR}). One is thus tempted to welcome the latter conclusion and drop the former, and simply ignore the perturbative results. In particular, the strong coupling problem we isolated in sect.~\ref{processes} might not be there---being associated with excitations that might  themselves not be there. Also, in the $\vec x(\vec \phi, t)$ parameterization of sect.~\ref{IR}, the vortex dynamics are essentially free---the only thing resembling an interaction is the volume-preserving constraint. It is not clear what `strong coupling' would mean in such a description of the system.

However, there is a number of confusing aspects that suggest that this optimistic attitude may be naive. The first is that even for theories where Coleman's theorem applies, one can recover correct information about physical quantities---like the spectrum for instance---by doing perturbative computations about the {\em wrong} semiclassical vacuum \cite{witten} (see also a related discussion in \cite{DS}). We do not know yet whether, and to what extent, the results of \cite{witten} apply in our case. If they do, they would demote our Coleman theorem-like result to a somewhat formal statement about the vacuum structure in the far infrared, with less crucial consequences for more local physics.

The second confusing fact---which points in the same general direction as the previous one---is that  for tiny but finite $c_T$ we have  a well defined effective theory that is perturbative up to some finite energy/momentum scale. For this effective theory there is no funny Coleman theorem-like behavior. All the symmetries that are spontaneously broken classically, remain so quantum-mechanically. The associated Goldstone bosons are described precisely by our perturbative analysis of sect.~\ref{processes}. On the other hand, as we stressed above, if $c_T \ll c_S$ we expect this theory to be physically equivalent to the $c_T = 0$ one for local questions, which do not rely crucially on the precise nature of the asymptotic states of the theory.
While this kind of logic may be misleading for massive gauge theories or massive gravity, where having or not having the mass really determines the number  of {\em local} (as opposed to asymptotic) physical degrees of freedom, here there is no such subtlety. The transverse degrees of freedom are perfectly physical even for vanishing $c_T$; they have  non-vanishing
conjugate momenta and are thus standard Hamiltonian degrees of freedom. Only, they do not feature wave solutions. 

It thus seems that for local questions the perturbative analysis of sect.~\ref{processes} should be perfectly fine. Of course the $S$-matrix is not a local quantity, and it may well be that the pathologies we encountered there are irrelevant for local questions. But we find it unlikely: if an effective theory exhibits strong coupling in scattering processes at some energy and momentum scales, it is probably useless for computing local correlation functions at the corresponding length scales. 
To settle the question one should compute directly local $n$-point functions, and see whether the perturbative expansion breaks down there. However, for the $S$-matrix we have a very powerful property---unitarity---that makes the tree level sufficient for such a question.
For local correlation functions instead, one should really ascertain the validity of the perturbative expansion by computing loop corrections---which, given that Lorentz invariance is spontaneously broken, is certainly doable but somewhat less transparent than usual. We leave this to future work.

Our results invite one to focus on correlators of quantities that are invariant under volume-preserving diffs. For instance, for a free massless scalar in $1+1$ dimensions, even though Coleman theorem applies, correlators of shift-invariant quantities are perfectly well-defined, and match what one would naively expect by doing perturbation theory about the wrong classical vacuum where the shift symmetry is spontaneously broken. Whether this property survives the inclusion of interactions and the generalization to our fluid, we do not know. If all diff-invariant correlators are well defined for our fluid, then one could in principle decide that those are the only observable quantities. This would correspond to {\em gauging} the problematic volume-preserving symmetry.  This is certainly an interesting possibility to consider, and maybe the theory defined this way would be consistent. Yet it is not clear to us what resemblance it would bear with a physical fluid: as we tried to make clear, at the classical level the volume preserving diffs are not a gauge redundancy---they are real symmetries acting on physical and measurable degrees of freedom.

On a different note, even if the effective theory is not strongly coupled, we find it interesting that microscopic vortices may impair its unitarity (sect.~\ref{viscosity}). We plan to make this statement more systematic and quantitative. In particular, it would be interesting to understand to what extent this effect can be parameterized as a new contribution to viscosity, and whether it has any relation to the conjectured viscosity-over-entropy bound \cite{KSS}.

\section*{Aknowledgments}

We would like to thank Nima Arkani-Hamed, Gregory Gabadadze, Dmitri Ivanov,  Eduardo Ponton, Slava Rychkov, Sergey Si\-bi\-ry\-a\-kov, and Dam Thanh Son for illuminating discussions. 
We also thank Georgios Pastras and Alessandro Vichi for collaboration in the early stages of this project.
The work of S.E.~is supported by the National Science Foundation through a Graduate Research Fellowship.
The work of A.N.~and J.W.~is supported by the DOE (DE-FG02-92-ER40699).
A.N.~would also like to thank the Institute of Theoretical Physics of the Ecole Polytechnique
F«ed«erale de Lausanne for hospitality during this project. The work of R.R. is supported by the Swiss National Science Foundation under contract 200021-125237. R.R.~ also acknowledges the hospitality of the Physics Department at Columbia University during the completion of this project.

\appendix

\section{Expanding the Lagrangian}

To carry out the expansion of the Lagrangian in fluctuations, eq.~(\ref{f}) is a slightly more convenient starting point than eq.~(\ref{action}).
More importantly, it is particularly convenient  to `pull out' the time derivatives from $B$:
\bea
B & = & \det{\di_\mu \phi_I \di^\mu \phi_J} \nonumber \\
&= & \det \big( \di \phi^T \cdot \di \phi - \dot {\vec \phi} \otimes \dot {\vec \phi} \big) \nonumber \\
& = & \det \Big(  \di \phi^T \cdot \big(1 - ((\di \phi^T)^{-1} \cdot \dot {\vec \phi}) \otimes  ((\di \phi^T)^{-1} \cdot \dot {\vec \phi})  \big) \cdot  \di \phi\Big) \nonumber \\
& = & (\det \di \phi)^2 \, \det\big(1 -\vec v \otimes \vec v \big) 
\label{B} \;,
\eea
where we defined the matrix $(\di \phi)_{ij} \equiv \di_i \phi_j$, and the vector
\be
\vec v \equiv (\di \phi^T)^{-1} \cdot \dot {\vec \phi}
\ee
Notice that $\vec v$ is the usual fluid velocity field---hence the name.
The last term in (\ref{B}) is easy to compute, for instance by going to a basis where, locally, the $x$ axis is aligned with $\vec v$. We get
\be
B = (\det \di \phi)^2 \big(1 - |\vec v|^2 ) 
\ee
and therefore
\be \label{expandL}
{\cal L} =  - w_0 f \big(\det \di \phi \,  \sqrt{1 - |\vec v|^2} \big)
\ee
Notice that we are essentially reproducing eq.~(88) of \cite{DGNR}. We now have to expand $f$ in powers of its argument, and its argument in powers of $\pi$.
The benefit of doing the expansion this way is that one now only has to expand in $\pi$ functions of $\di \phi = 1 + \di \pi$, rather than of $\di_\mu \phi \di^\mu \phi = 1 + \di \pi + \di \pi^T + \di \pi^T \di \pi + \dot \pi \dot \pi$. This simplifies the algebra considerably. We just need
\bea
(\di \phi)^{-1} & \simeq & 1 - \di \pi + \di \pi^2 \\
\det \di \phi & = & 1 + [\di \pi] + \sfrac12 \big( [\di \pi]^2 - [\di \pi^2]\big)
+ \sfrac16 \big( [\di \pi]^3 - 3 [\di \pi] [\di \pi^2] + 2 [\di \pi^3] \big)
\eea 
(the determinant of an $n \times n$ matrix stops at $n$-th order.)
A straightforward Taylor-expansion of (\ref{expandL}) up to fourth order then yields eq.~(\ref{full_action}).

It is worth pointing out two  sources of non-trivial cancellations, with important physical consequences. The first forces all Lagrangian terms that do not involve time-derivatives to be weighed by $c_s^2 = f''(1)$ or by higher derivatives of $f$, as manifest in eq.~(\ref{full_action}). That is, no such term is coming from the expansion of (\ref{expandL}) at first order in $f$'s argument, i.e.~with a coefficient $f'(1) = 1$. The reason is simple: neglecting time-derivatives, the term proportional to $f'(1)$ would be
\be
{\cal L} \supset  - w_0  f'(1) \big( \det \di \phi -1\big) \; ,
\ee
which is a total derivative:
\be \label{total_derivative}
 \det \di \phi = \epsilon \, \epsilon \, \di \phi \, \di \phi \, \di \phi = \di \big(\epsilon \, \epsilon \, \phi \, \di \phi \, \di \phi  \big)
\ee
Since one expects higher derivatives of $f$ to be naturally of order $c_s^2$, for a non-relativistic fluid this cancellation has the effect of weakening the interactions considerably, or equivalently of raising the strong-coupling scale compared to what one may have naively guessed before carrying out the expansion.

The second cancellation involves the transverse phonons only, and has also the effect of weakening some interactions and correspodingly raising the vortex strong-coupling scale.
Consider an interaction term with spatial derivatives only, and assume that at least one of the phonons entering the corresponding vertex is transverse. As {\em not} manifest from the Lagrangian (\ref{full_action}), such a vertex yields zero. The reason is that we can perform a non-linear field redefinition that makes vortices disappear from all Lagrangian terms without time-derivatives.  The trick is to define $\vec \pi$ so that
\be
\det \di \phi = 1 + \vec \nabla \cdot \vec \pi
\ee
{\em exactly}. That this is possible follows from eq.~(\ref{total_derivative})---we may as well call the total derivative on the r.h.s.~$1 + \vec \nabla \cdot \vec \pi$.
This matches our original definition of $\vec \pi$ at linear order, and as a consequence it does not affect  the $S$-matrix. But now it is clear that vortex interactions will only come from the $|\vec v|^2$ part  of (\ref{expandL}), and will thus involve at least two time-derivatives. The downside is that in these variables the structure of the Lagrangian will be more complicated than eq.~(\ref{full_action}); in particular we will not have exactly one derivative acting on each field. For this reason we stick to the original definition of the phonon field and to eq.~(\ref{full_action}), but we should expect non-trivial cancellations when computing $S$-matrix elements involving vortices, as we indeed find in sect.~\ref{processes}.


\section{The $S$-matrix, cross-sections, and decay rates}
Here we briefly review the standard relativistic formulae for the $S$-matrix and related physical quantities like cross sections and decay rates, and derive the modifications needed for applying them to our  $c \neq 1$ case. The rules we will derive are straightforwardly generalizable to the case of different fields with different propagation speeds. 

We will borrow the conventions of Peskin-Shroeder \cite{Peskin}. In particular, we use the so-called relativistic normalization for one-particle states:
\be
\langle \vec p \, | \vec q \, \rangle = (2 E) \, (2\pi)^3 \delta^{3}(\vec p- \vec q) \; ,
\ee
(we are suppressing spin labels---their inclusion is straightforward)
and of course the vacuum state $| 0 \rangle$ is normalized to one.
This way, a relativistic canonically normalized scalar field $\phi(x)$ obeys
\be \label{canonical}
\langle 0 | \phi(x) | \vec p \, \rangle = e^{-i (E t - \vec p \cdot \vec x)} \; ,
\ee
and consequently the momentum-space  Feynman rules assign a wavefunction one to external spin-0 states.
One thus has that for a $2 \to n_f$ scattering process the infinitesimal cross section is
\be \label{sigma}
d \sigma = \frac{1}{2 E_A} \frac{1}{2 E_B}  \frac{1}{|v_A - v_B|} \, \big| {\cal M}_{AB \to f} \big|^2 \, d \Pi_{n_f} \; .
\ee
Here ${\cal M}_{i \to f}$ is the amplitude computed according to the standard relativistic Feynman rules, and defined by
\be \label{M}
\langle \vec q_1 \dots \vec q_{n_f} \, | (S-1)  | \vec p_1 \dots \vec p_{n_i} \rangle = (2\pi)^4 \, \delta^3({\rm momentum}) \, \delta({\rm energy}) \cdot {i {\cal M}_{i \to f}} \;,
\ee
and the $d \Pi_{n_f}$ is the relativistic final-state phase-space:
\be \label{phase_space}
d \Pi_{n_f} = (2\pi)^4 \, \delta^3({\rm momentum}) \, \delta({\rm energy}) \cdot \bigg( \prod_f \frac{d^3 q_f}{(2 \pi)^3} \frac{1}{2 E_f} \bigg)
\ee
Finally, $|v_A - v_B|$ is the relative velocity between the two colliding beams as measured in the lab frame.
Likewise for a $1 \to n_f$ decay process, the infinitesimal rate is
\be \label{decay rate}
d \Gamma = \frac{1}{2 E_A} \, \big| {\cal M}_{A \to f} \big|^2 \, d \Pi_{n_f} \; .
\ee

First, let us check the dimensions of these quantities, by keeping $\hbar$ dimensionless but the speed of light dimensionful. That is, let's give ${\rm energy=1/time}$ and ${\rm momentum=1/length}$ different units. From their definitions, eqs.~(\ref{M}, \ref{phase_space}), for the amplitude and phase-space element we get
\be
\big[ {\cal M}\big] = E k^3 \big( E/k^3 \big) ^{\frac{n_i+n_f}{2}}  \; , \qquad  \big[ d \Pi \big] = (E k^3)^{-1} \big( E/k^3\big)^{-n_f} \; .
\ee
The cross-section and decay rate thus have dimensions
\be
\big[d \sigma \big] = k^{-2} = {\rm area} \; , \qquad  \big[ d \Gamma \big] = E = 1/{\rm time} \; ,
\ee
as they should. This means that the above fomulae are already dimensionally correct with no need of explicit powers of the speed of light.

Next, we notice that nowhere is Lorentz invariance assumed in deriving the Feynman rules and the above expressions for $\sigma$ and $\Gamma$. This is evident e.g.~in the derivation of ref.~\cite{Peskin}, apart from the relative velocity factor in the cross section. However that too is independent of Lorentz invariance, for it arises from the integral of an energy delta-function over the longitudinal (w.r.t.~to the collision direction) momenta of the incoming wave-packets:
\bea
\int dk^z_A \, dk^z_B  \, \delta(k^z_A + k^z_B - P^z_f) \, \delta(E_A + E_B - E_f) & = & \int dk^z_A  \, \delta(E_A + E_B - E_f) \big|_{k^z_B = P^z_f - k^z_A} \\
& = & \Big| \frac{\di E_A}{d k^z_A} - \frac{\di E_B}{\di k^z_B}\Big|^{-1} \; .
\eea                                                      
For each wave-packet, the derivative of the energy w.r.t.~the corresponding momentum is the wave-packet's group-velocity, independently of the actual form of the dispersion law $E(k)$. The above thus yields the factor $1/|v_A - v_B|$ in the cross section, regardless of Lorentz-invariance.

The bottom line is, much ado about nothing. We can use the standard relativistic Feynman rules and formulae for infinitesimal cross-sections and rates for our  non-relativistic case as well, with no modifications, even when  different fields have different speeds. The only subtlety we should keep in mind is that canonically normalized fields obey eq.~(\ref{canonical}), times possible polarization factors for non-scalar particles. This means that a scalar field $\phi$ thus normalized should appear in the action as
\be
S = \int d^3x dt \, \sfrac12 \dot \phi^2 + \dots  \; ,
\ee
so that single-particle states are eigenstates of the free Hamiltonian with the right energy:
\be
w_0 | \vec p \, \rangle = \Big( \int d^3x \, \sfrac12 \dot \phi^2 + \dots \Big)  | \vec p \, \rangle =  E(\vec p \,)  | \vec p \, \rangle \; .
\ee

As a check that these conclusions make sense, we estimate the cross section for sound wave-sound wave elastic scattering and show that, indeed, we have strong-coupling at the correct energy. From Feynman rules applied to the Lagrangian (\ref{full_action}) we have
\be
{\cal M} \sim c_s^2 \frac{k^4}{w_0}
\ee
where the factor of $w_0$ comes from the non-canonical normalization of $\pi^I$. The final state phase space (\ref{phase_space}) is of order
\be
\Pi_f \sim \frac{1}{k^3} \frac{1}{E} \big( k^3/ E \big)^2 \; ,
\ee
and the relative velocity is of course $2c_s$, so that the cross-section (\ref{sigma}) is
\be
\sigma \sim \frac{1}{k^2} \bigg( \frac{k^4}{w_0 c_s} \bigg)^2 \; .
\ee
This agrees with our estimates of sect.~\ref{strong_coupling}---see the last paragraph of sect.~\ref{LLtoLL}.


\section{Phase space}

We are mostly interested in a two-particle final state, possibly with two independent propagation speeds. The infinitesimal phase space is
\be
d \Pi_2 = (2\pi)^4 \, \delta^3(\vec P - \vec q_1 - \vec q_2) \, \delta(E - E_1 - E_2) \cdot \frac{d^3 q_1}{(2 \pi)^3}\frac{d^3 q_2}{(2 \pi)^3}  \frac{1}{2 E_1}  \frac{1}{2 E_2} \; ,
\ee
where $E$ and $\vec P$ are the total energy and momentum.
The integral in $\vec q_2$ eliminates the momentum-conservation delta-function. Then we are left with
\be
d \Pi_2 = \frac{d \Omega}{(2 \pi)^2} \, q_1^2 d q_1 \frac{1}{2 E_1 \, 2 E_2} \delta(E - E_1 - E_2)  \; , 
\ee
with the understanding that $E_2$ be evaluated at $\vec q_2 = \vec P - \vec q_1$. We have
\be
\delta(E - E_1 - E_2) = \frac{\delta(q_1 - \bar q_1)}{\big|  \frac{\di E_1}{\di q_1} + \frac{\di E_2}{\di q_2} \frac{\di q_2}{\di q_1}  \big|}
\ee
and
\be
\frac{\di q_2}{\di q_1} \equiv \frac{\di \big |\vec P  - \vec q_1 |}{\di q_1} = \frac{q_1- P \cos \theta}{q_2} \; ,
\ee
where $\theta$ is the angle between $\vec q_1$ and $\vec P$. On the other hand, the derivatives of the energies w.r.t.~the corresponding momenta are the particles' group velocities. Integrating over $q_1$ we thus get
\be
d \Pi_2 = \frac{d \Omega}{16 \pi^2} \frac{q_1^2 q_2}{E_1 E_2} \frac{1}{\big| c_1 q_2 + c_2 q_1 - c_2 P \cos \theta \big|}
\ee
For a linear dispersion law like in our case, $E_a = c_a q_a $, we finally have
\be \label{non-zero momentum 2 particle phase space}
d \Pi_2 = \frac{d \Omega}{16 \pi^2} \frac{1}{c_1 c_2} \frac{q_1}{\big| c_1 q_2 + c_2 q_1 - c_2 P \cos \theta \big|} \; .
\ee
In special circumstances there are further simplifications:
\begin{itemize}
\item[{\em i)}] 
For scattering processes at zero total momentum, we can set $P=0$ and $q_1 = q_2$. We get
\be \label{dPi_zeroP}
d \Pi_2 = \frac{d \Omega}{16 \pi^2} \cdot \frac{1}{c_1 c_2(c_1 + c_2)} \qquad (\vec P=0)\; .
\ee
\item[{\em ii)}] 
For decay processes at finite total $\vec P$, but when one of the final particles is much slower that the other, barring an hierarchy between $q_1$ and $q_2$ we have
\be
d \Pi_2 \simeq \frac{d \Omega}{16 \pi^2} \, \frac{1}{c_1^2 c_2} \, \frac{q_1}{q_2}  \qquad (c_2 \ll c_1) \; .
\ee
Of course the ratio $q_1/q_2$ depends non-trivially on the angle $\theta$ we are supposed to integrate over---which we can take to be the angle between $\vec q_1$ and $\vec P$. We have:
\be
\frac{q_1}{q_2} \simeq \frac{1}{2 \sin \theta/2} \qquad (c_2 \ll c_1  , \; \vec P \neq 0) \; .
\ee
Overall we thus get
\be	 \label{dPi_nonzeroP}
d \Pi_2 \simeq \frac{d \Omega}{32 \pi^2} \, \frac{1}{c_1^2 c_2} \frac{1}{\sin \theta/2} \qquad (c_2 \ll c_1  , \; \vec P \neq 0) \; .
\ee
Notice that this is regular at $\theta = 0$, thus making our `barring an hierarchy \dots' approximation under control. That is, eq.~(\ref{dPi_nonzeroP}) is the correct phase-space element at lowest order in $c_2/c_1$.
\end{itemize}



\begin{thebibliography}{99}
\small

\bibitem{LL}
  E.~M.~Lifshitz and L.~P.~Pitaevskii,
  ``Statistical physics. Part 2: Theory of the condensed state,''
{\it  Oxford, UK: Butterworth-Heinemann (1980) 387p}



\bibitem{NS}
  D.~Nickel and D.~T.~Son,
  ``Deconstructing holographic liquids,''
  arXiv:1009.3094 [hep-th].


\bibitem{EJL}
  M.~Edalati, J.~I.~Jottar and R.~G.~Leigh,
  ``Transport coefficients at zero temperature from extremal black holes,''
  JHEP {\bf 1001}, 018 (2010)
  [arXiv:0910.0645 [hep-th]].

\bibitem{Weinberg}
S.~Weinberg, ``Gravitation and Cosmology," {\it John Wiley \& Sons (1972) 657p.}


\bibitem{DGNR}
  S.~Dubovsky, T.~Gregoire, A.~Nicolis and R.~Rattazzi,
  ``Null energy condition and superluminal propagation,''
  JHEP {\bf 0603}, 025 (2006)
  [arXiv:hep-th/0512260].


\bibitem{ghost}
  N.~Arkani-Hamed, H.~C.~Cheng, M.~A.~Luty and S.~Mukohyama,
  ``Ghost condensation and a consistent infrared modification of gravity,''
  JHEP {\bf 0405}, 074 (2004)
  [arXiv:hep-th/0312099].



\bibitem{Peskin}
  M.~E.~Peskin and D.~V.~Schroeder,
  ``An introduction to quantum field theory,''
{\it  Reading, USA: Addison-Wesley (1995) 842 p}

\bibitem{coleman}
  S.~R.~Coleman,
  ``There are no Goldstone bosons in two-dimensions,''
  Commun.\ Math.\ Phys.\  {\bf 31}, 259 (1973).

\bibitem{witten}
  E.~Witten,
  ``Chiral symmetry, the $1/N$ expansion, and the $SU(N)$ Thirring model,''
  Nucl.\ Phys.\  B {\bf 145}, 110 (1978).


\bibitem{DS}
  S.~Dubovsky and S.~Sibiryakov,
  ``Superluminal travel made possible (in two dimensions),''
  JHEP {\bf 0812}, 092 (2008)
  [arXiv:0806.1534 [hep-th]].

\bibitem{KSS}
  P.~Kovtun, D.~T.~Son and A.~O.~Starinets,
  ``Viscosity in strongly interacting quantum field theories from black hole
  physics,''
  Phys.\ Rev.\ Lett.\  {\bf 94}, 111601 (2005)
  [arXiv:hep-th/0405231].




\end{thebibliography}
\end{document}